\RequirePackage{fix-cm}
\documentclass[twocolumn,epjc3]{svjour3}
\smartqed  % flush right qed marks, e.g. at end of proof
\journalname{Eur. Phys. J. A}
\usepackage{amsmath}
\usepackage{graphicx}
\usepackage{amsfonts}
\usepackage{amssymb}
\usepackage{graphicx}
\usepackage{color}
\usepackage{lipsum}
\usepackage{multirow}
\usepackage{widetext}
\usepackage{flushend}
\usepackage{cuted}

\begin{document}

\title{Study on relativistic transformations for thermodynamic quantities: Boltzmann-Gibbs and Tsallis blast-wave models}



\author{A.S.~Parvan\thanksref{e1,addr1,addr2}
}

\thankstext{e1}{e-mail: parvan@theory.nipne.ro}


%\institute{Bogoliubov Laboratory of Theoretical Physics, Joint Institute for Nuclear Research, Dubna, 141980, Russia \label{addr1}
%           Department of Theoretical Physics, Horia Hulubei National Institute for R$\&$D in Physics and Nuclear Engineering, M\u{a}gurele, Ilfov, 077125, Romania \label{addr2}
%}

\institute{BLTP, Joint Institute for Nuclear Research, Dubna, 141980, Russia \label{addr1}
           \and
           DTP, Horia Hulubei National Institute for R$\&$D in Physics and Nuclear Engineering, M\u{a}gurele, Ilfov, 077125, Romania \label{addr2}
}

\date{Received: date / Accepted: date}

\maketitle

\begin{abstract}
This study derives the relativistic transformations of thermodynamic quantities from the Lorentz transformations applied to the four-momentum components of a thermodynamic system, which is stationary in the inertial reference frame $ K_0 $ and moves at constant velocity relative to the laboratory frame $ K $. Thermodynamic variables are introduced into the formalism via the zeroth component of the four-momentum in $ K_0 $, representing the system's internal energy. Entropy and particle number are relativistic invariants, while the volume undergoes Lorentz contraction. By treating the three-momentum as an independent state variable, thermodynamic quantities are defined by differentiating the zeroth component of the four-momentum (the Hamiltonian) in the reference frame $ K $ with respect to the independent state variables, yielding the fundamental thermodynamic potential. This approach results in the Non-Planck transformations, which differ from the Planck transformations by a factor of $ \alpha $. In contrast, by adopting the three-velocity as an independent state variable, thermodynamic quantities are obtained by differentiating the negative Lagrangian, derived from the zeroth component of the four-momentum via Legendre transformations, with respect to the independent state variables, producing the conjugate fundamental thermodynamic potential. This yields the Planck transformations. Conversely, the Ott transformations are derived from the zeroth component of the four-momentum by treating velocity as an independent state variable. This approach conflicts with the principles of mechanics, resulting in an energy that does not qualify as a thermodynamic potential. To validate these findings, we analyze an ultrarelativistic ideal gas of quarks and gluons within the Stefan-Boltzmann limit. Furthermore, we develop consistent Boltzmann-Gibbs and Tsallis blast-wave models for finite-volume freeze-out firecylinders in heavy ion collisions, incorporating Planck and Ott transformations. Comparative analysis demonstrates that Planck transformations yield consistent transverse momentum distributions of hadrons, whereas Ott transformations result in discrepancies.
%\keywords{Relativistic thermodynamics \and Lorentz transformations \and Planck's formalism \and Ott's formalism \and Quark-Gluon Plasma \and Blast-Wave Model}
\end{abstract}

\section{Introduction}\label{intro}
The interplay between special relativity and thermodynamics forms a cornerstone of modern physics, particularly when analyzing systems at relativistic speeds. Thermodynamic quantities—such as energy, entropy, and temperature—must be carefully reformulated within relativistic frameworks to ensure consistency with the theory of special relativity. The challenge of establishing a robust foundation for relativistic thermodynamics in moving systems remains unresolved, as noted in Refs.~\cite{Haar,Farias,Nakamura}. Central to this reformulation are the Planck and Ott transformations, two distinct approaches to describing how thermodynamic quantities transform between inertial reference frames. The Planck transformation, developed by Mosengeil, Planck, and Einstein~\cite{Mosengeil,Planck,Einstein}, is grounded in preserving thermodynamic relations, while the Ott transformation, proposed by Ott~\cite{Ott}, provides an alternative relativistic interpretation. These approaches offer competing yet complementary perspectives. According to Planck’s theory, the temperature $T$ in an inertial reference frame moving at a constant velocity $v$ relative to a thermodynamic system with a rest temperature $T_{0}$ is given by $T=T_{0}/\gamma<T_{0}$, where $\gamma = (1-v^{2}/c^{2})^{-1/2}$. Conversely, Ott’s formalism suggests $T=\gamma T_{0}>T_{0}$. This fundamental disagreement, alongside other proposed frameworks, has fueled significant debate in relativistic thermodynamics, as evidenced by works in Refs.~\cite{Arzelies,Kibble,Sutcliffe,Landsberg,Cavalleri,Newburgh,Yuen,Callen,Treder,Pathria,Fenech,Kampen,Israel,Landsberg1996,Mares,Papadatos,Parga,Heras,Mares2017,Mi,Agmon,Balescu,Guemez,Gavassino,Przanowski,Montakhab,Biro,Hao,Parvan2019}.
These frameworks are critical for applications in high-energy physics, astrophysics, and cosmology, where systems like relativistic gases, neutron stars, or quark-gluon plasmas operate under extreme conditions. Relativistic thermodynamics, developed from the perspective of special relativistic hydrodynamics, is discussed in Refs.~\cite{Biro,Biro1}. Studies on the application of four-temperature~\cite{Haar} in statistical mechanics are explored in Refs.~\cite{Becattini1,Becattini2}. The covariant formulation of thermodynamics was explored in Gavassino's work~\cite{Gavassino22}.

This paper presents a novel method for deriving the Planck and Ott transformations, grounded in the framework of thermodynamic potentials, which forms the foundation of equilibrium thermodynamics. We examine a thermodynamic system at rest in an inertial reference frame $K_0$, moving at constant velocity relative to the laboratory frame $K$. By integrating thermodynamic potentials with relativistic dynamics, we derive transformations for key thermodynamic quantities using the fundamental thermodynamic potential—total relativistic energy (the Hamiltonian, expressed as a function of momentum as the independent state variable) — and its conjugate potential (the negative Lagrangian, expressed as a function of velocity as the independent state variable). These potentials serve as the foundation for deriving other potentials and ensembles through Legendre transformations or entropy substitution, with the system's rest energy (internal energy) serving as the fundamental thermodynamic potential in $K_0$. Our approach ensures consistency with the first law of thermodynamics (fundamental equation of thermodynamics) and the principle of entropy invariance. The fundamental thermodynamic potential, represented by the Hamiltonian (a function of momentum), yields relativistic transformations that deviate from the Planck transformations by a factor of $\alpha$. Conversely, the conjugate potential, defined as the negative Lagrangian, exactly reproduces the Planck transformations. In contrast, the Ott transformations are derived from the total relativistic energy, expressed as a function of velocity as an independent state variable. However, using total energy in this manner does not correspond to a thermodynamic potential. This paper elucidates the origins of the Planck and Ott transformations. Note that the Planck transformation for temperature, derived from considerations similar to those in our paper, was proposed by Agmon in Ref.~\cite{Agmon}, where the negative Lagrangian is referred to as the "inner energy" or "unordered part" of energy~\cite{Pathria}.

We systematically investigate relativistic transformations using an ultrarelativistic ideal gas of quarks and gluons in the Stefan-Boltzmann limit, analyzing the behavior of temperature and pressure in a relativistically moving quark-gluon plasma system. Additionally, we develop Boltzmann-Gibbs and Tsallis blast-wave models within a finite freeze-out firecylinder volume, incorporating Planck and Ott transformations to study transverse momentum distributions in the laboratory reference frame $K$. By comparing global equilibrium transverse momentum distributions of hadrons, derived from conventional Boltzmann-Gibbs and Tsallis formalisms, with local equilibrium distributions obtained from the blast-wave models, we demonstrate that the Planck transformations yield consistent and accurate results, whereas the Ott transformations introduce significant discrepancies.

The paper is structured as follows: Section 2 introduces the definition of the relativistic thermodynamic system. Section 3 focuses on deriving relativistic transformations for thermodynamic quantities. In Section 4, we explore the properties of a moving quark-gluon plasma (QGP) system. Section 5 presents the development of the Boltzmann-Gibbs and Tsallis blast-wave models. The final section provides discussions and conclusions.

%\section{Relativistic thermodynamic systems: The fundamental ensemble}\label{sec2}
\section{Relativistic thermodynamic systems}\label{sec2}
\subsection{Relativistic Mechanics}
Consider a thermodynamic system $\mathcal{A}$ with mass $M$ moving at relativistic speeds, observed in two inertial reference frames, $K$ and $K_0$. The reference frame $K_0$ moves in a straight line at a constant velocity $\mathbf{v}=(v, 0, 0)$ relative to the stationary laboratory frame $K$, aligned along the positive directions of the $x$ and $x_0$ axes. In the frame $K_0$, the 4-momentum of the system $\mathcal{A}$ is given by $P_0^{\mu}=(E_0, \mathbf{P}_0)$, while in the frame $K$, it is $P^{\mu}=(E, \mathbf{P})$, where $\mathbf{P}_0$ and $\mathbf{P}$ represent the 3-momenta, and $E_0$ and $E$ denote the total energies of the system $\mathcal{A}$. The Lorentz transformations relating the 4-momenta of the system $\mathcal{A}$ between these frames can be expressed as~\cite{Landau}
\begin{align}\label{3}
  E &= \gamma \left(E_{0}+v P_{0x}\right),  \quad  P_{x} = \gamma \left(P_{0x}+v E_{0}\right), \\ \label{3a}
  E_{0} &= \gamma \left(E-v P_{x}\right) = u_{\mu}P^{\mu}, \quad  P_{0x} = \gamma \left(P_{x}-v E\right).
\end{align}
The $y$ and $z$ components of the 3-momentum are conserved, such that $P_{y} = P_{0y}$ and $P_{z} = P_{0z}$, respectively. In this context, $u^{\mu} = \gamma (1, \mathbf{v})$ denotes the 4-velocity, with $\gamma$ being the Lorentz factor:
\begin{equation}\label{5}
  \gamma = \frac{1}{\sqrt{1-\mathbf{v}^{2}}}.
\end{equation}
Here and throughout this paper we use the system of natural units $\hbar=c=k_{B}=1$.

In this paper, we assume that the thermodynamic system $\mathcal{A}$ is stationary within the moving reference frame $K_0$. Consequently, $\mathbf{P}_0 = 0$, and the 4-momentum simplifies to $P_0^{\mu} = (E_0, 0)$, where the total energy is $E_0 = M$. By substituting this value of $\mathbf{P}_0$ into Eq.~(\ref{3}), we obtain the Lorentz transformations for total energy and momentum of thermodynamic system as~\cite{Landau}
\begin{equation}\label{6}
  E = \gamma E_{0}, \qquad   \mathbf{P} = E \mathbf{v}  = \gamma E_{0} \mathbf{v},
\end{equation}
where $\mathbf{P}=(P_{x},0,0)$. In this scenario, the 3-velocity of the thermodynamic system $\mathcal{A}$ corresponds to the velocity $\mathbf{v}$ of the reference frame $K_0$, as expressed in~\cite{Landau}:
\begin{equation}\label{8}
  \mathbf{v} = \frac{\mathbf{P}}{E}.
\end{equation}
Substituting Eq.~(\ref{8}) into Eq.~(\ref{5}) and using Eq.~(\ref{6}), we obtain the relativistic dispersion relation, as given in Ref.~\cite{Landau}:
\begin{equation}\label{8a}
  E=\sqrt{\mathbf{P}^{2}+E_{0}^{2}},
\end{equation}
where $E$ is the total energy, $\mathbf{P}$ denotes the three - momentum vector, and $E_0$ represents the rest energy. This relation can also be derived from the invariance of the four-momentum squared, $P_\mu P^\mu = E_0^2$. Here, the total energy $E$ corresponds to the relativistic Hamiltonian for a freely moving particle without interactions with three-momentum $\mathbf{P}$ and rest energy $E_0$ (see Eq.~(9.7) in Ref.~\cite{Landau}). The energy expressed as a function of momentum is called the Hamiltonian function~\cite{Landau}.

The 3-velocity $\mathbf{v}$ and 3-momentum $\mathbf{P}$ of the thermodynamic system $\mathcal{A}$ in the reference frame $K$ are interconnected through Eq.~(\ref{8}). These quantities are conjugate variables. Specifically, if the 3-momentum $\mathbf{P}$ is treated as an independent state variable, the 3-velocity $\mathbf{v}$ becomes a dependent variable that can be derived from it, and vice versa. This duality gives rise to two distinct approaches in mechanics. In the first case, when the 3-momentum $\mathbf{P}$ is the independent state variable, it aligns with the Hamiltonian function $E$ within the framework of Hamiltonian mechanics, and the velocity $\mathbf{v}$ is expressed as a function of $\mathbf{P}$. In the second case, when the 3-velocity $\mathbf{v}$ is the independent state variable, it corresponds to the Lagrangian function $L$ within Lagrangian mechanics, and the momentum $\mathbf{P}$ is determined as a function of $\mathbf{v}$. The Lagrangian $L$ of the thermodynamic system $\mathcal{A}$ in the reference frame $K$ is obtained from the Hamiltonian $E$ through a Legendre transform, as described in Ref.~\cite{Landau} (cf. Eq.~(\ref{3a})):
\begin{equation}\label{9}
  L = \mathbf{v} \cdot \mathbf{P} - E =-\frac{E_{0}}{\gamma} = - \frac{u_{\mu}P^{\mu}}{\gamma},
\end{equation}
where $\mathbf{v}$ is the three-velocity, $\mathbf{P}$ is the three-momentum, $E$ is the total energy, $E_0$ is the rest energy, and $\gamma = (1 - \mathbf{v}^2)^{-1/2}$ is the Lorentz factor. Alternatively, the Lagrangian $L$ can be derived from the action for a single particle with three-velocity $\mathbf{v}$ and rest energy $E_0$, as presented in Eq.~(8.2) of Ref.~\cite{Landau}.

\subsection{Equilibrium thermodynamics}
In equilibrium thermodynamics, a thermodynamic system is defined by its boundary conditions, which isolate it from the surrounding environment, and by the exchange of energy (heat and work) and mass across those boundaries~\cite{Prigogine1,Kvasnikov}. The thermodynamic state of the system is determined by independent state variables, selected based on these boundary conditions. All other thermodynamic quantities, together with the state variables and the corresponding thermodynamic potential, form a complete set of variables characterizing the system and are considered dependent variables. The chosen independent state variables and their associated thermodynamic potential define the system's ensemble. Once selected, these independent state variables cannot be arbitrarily changed during problem-solving, and any transition to a different set must strictly follow the relevant mathematical rules~\cite{Prigogine2}. A thermodynamic system $\mathcal{A}$ is fully described by its fundamental thermodynamic potential $\mathcal{F}$ (energy), expressed as a function of its state variables. The thermodynamic quantities of the system $\mathcal{A}$ are derived from the first and second partial derivatives of the fundamental thermodynamic potential $\mathcal{F}$ with respect to these state variables~\cite{Parvan15}. Transitioning to another ensemble involves replacing the independent state variables and shifting to a different thermodynamic potential, either through Legendre transformations or by substituting the fundamental thermodynamic potential with one of its state variables, such as entropy.

In equilibrium thermodynamics, transitions between thermodynamic states occur via quasi-static processes, which can proceed in either direction with infinitesimal changes in the parameters governing the equilibrium~\cite{Prigogine1,Kvasnikov}.

\subsubsection{Motionless system in reference frame $K_0$}
The thermodynamic system $\mathcal{A}$ is at rest in the reference frame $K_0$, which is in motion relative to laboratory frame $K$. Below, we outline the primary ensembles of equilibrium thermodynamics for this system.
\paragraph{1. Fundamental ensemble $(S_0, V_0, N_0)$}
In equilibrium thermodynamics, the fundamental ensemble is characterized by the fundamental thermodynamic potential, which is the internal energy of the system. For a motionless system, the internal energy corresponds to the rest energy $E_0$ of the system (see Ref.~\cite{Nakamura}), expressed as a function of the thermodynamic state variables $(S_0, V_0, N_0)$ (cf. Eq.~(\ref{3a})):
\begin{equation}\label{9a}
  E_{0}=E_{0}(S_{0},V_{0},N_{0})=u^{\mu}P_{\mu},
\end{equation}
where $S_0$ represents the rest entropy, $V_0$ denotes the rest volume, and $N_0$ indicates the rest number of particles. The differential of the thermodynamic potential $E_0$ with respect to the state variables $(S_0, V_0, N_0)$ is given by~\cite{Parvan2019}:
\begin{align}\label{1}
   dE_{0} &=  T_{0}dS_{0} - p_{0}dV_{0} + \mu_{0} dN_{0}
   = \delta Q_{0} - \delta W_{0} \nonumber \\ & + \delta E_{0}^{mat}, \\ \label{2}
   T_{0}  &\equiv  \left(\frac{\partial E_{0}}{\partial S_{0}}\right)_{V_{0}N_{0}}, \quad  p_{0} \equiv - \left(\frac{\partial E_{0}}{\partial V_{0}}\right)_{S_{0}N_{0}}, \nonumber \\    \mu_{0} &\equiv  \left(\frac{\partial E_{0}}{\partial N_{0}}\right)_{S_{0}V_{0}},
\end{align}
where $T_0$, $p_0$, and $\mu_0$ represent the temperature, pressure, and chemical potential, respectively. Additionally, $\delta Q_0 = T_0 dS_0$ corresponds to the heat transfer, $\delta W_0 = p_0 dV_0$ denotes the mechanical work due to volume change, and $\delta E_0^{\text{mat}} = \mu_0 dN_0$ reflects the energy change due to matter exchange for the thermodynamic system in the reference frame $K_0$. Equation~(\ref{1}) represents \emph{the first law of thermodynamics (the law of conservation of energy)}, while Eq.~(\ref{2}) defines the thermodynamic quantities $T_0$, $p_0$, and $\mu_0$ in the reference frame $K_0$, where the thermodynamic system is at rest.

\paragraph{2. Canonical ensemble $(T_0, V_0, N_0)$}
The canonical ensemble is derived from the fundamental ensemble by replacing the variable $S_0$ with $T_0$. The thermodynamic potential for the canonical ensemble, the Helmholtz free energy $F_0$, is obtained from the fundamental thermodynamic potential $E_0$ through a Legendre transform:
\begin{equation}\label{ad1}
  F_{0}(T_0, V_0, N_0) = E_{0} - T_{0} S_{0},
\end{equation}
where $S_0(T_0, V_0, N_0)$ is determined by solving the equation $T_0 = \frac{\partial E_0}{\partial S_0}$ at constant $T_0$. The differential of the Helmholtz free energy $F_0$ with respect to the state variables $(T_0, V_0, N_0)$ is given by:
\begin{align}\label{ad2}
   dF_{0} &= -S_{0} dT_{0} - p_{0}dV_{0} + \mu_{0} dN_{0}, \\ \label{ad3}
   S_{0}  &=  -\left(\frac{\partial F_{0}}{\partial T_{0}}\right)_{V_{0}N_{0}}, \quad  p_{0} = - \left(\frac{\partial F_{0}}{\partial V_{0}}\right)_{T_{0}N_{0}}, \nonumber \\    \mu_{0} &= \left(\frac{\partial F_{0}}{\partial N_{0}}\right)_{T_{0}V_{0}},
\end{align}
where $S_0$ is the entropy, $p_0$ is the pressure, and $\mu_0$ is the chemical potential of the system.

\paragraph{3. Isobaric ensemble $(T_0, p_0, N_0)$}
The isobaric ensemble is derived from the fundamental ensemble by replacing the variables $S_0$ and $V_0$ with $T_0$ and $p_0$, respectively. The thermodynamic potential for the isobaric ensemble, known as the Gibbs free energy $G_0$, is obtained from the fundamental thermodynamic potential $E_0$ through a Legendre transform:
\begin{equation}\label{ad4}
  G_{0}(T_0, p_0, N_0) = E_{0} - T_{0} S_{0} + p_{0} V_{0},
\end{equation}
where $S_0(T_0, p_0, N_0)$ and $V_0(T_0, p_0, N_0)$ are determined by solving the equations $T_0 = \frac{\partial E_0}{\partial S_0}$ and $p_0 = -\frac{\partial E_0}{\partial V_0}$ at constant temperature $T_0$ and pressure $p_0$, respectively. The differential of the Gibbs free energy, known as the Gibbs relation, is given by:
\begin{align}\label{ad5}
   dG_{0} &= -S_{0} dT_{0} + V_{0}dp_{0} + \mu_{0} dN_{0}, \\ \label{ad6}
   S_{0}  &= -\left(\frac{\partial G_{0}}{\partial T_{0}}\right)_{p_{0}N_{0}}, \quad  V_{0} =  \left(\frac{\partial G_{0}}{\partial p_{0}}\right)_{T_{0}N_{0}}, \nonumber \\    \mu_{0} &=  \left(\frac{\partial G_{0}}{\partial N_{0}}\right)_{T_{0}p_{0}},
\end{align}
where $S_0$ represents the entropy, $V_0$ the volume, and $\mu_0$ the chemical potential of the system.

\paragraph{4. Grand canonical ensemble $(T_0, V_0, \mu_0)$}
\ The grand canonical ensemble is derived from the fundamental ensemble by replacing the variables $S_0$ and $N_0$ with $T_0$ and $\mu_0$, respectively.
The thermodynamic potential for the grand canonical ensemble, known as the grand potential $\Omega_0$, is obtained from the fundamental thermodynamic potential $E_0$ via a Legendre transform:
\begin{equation}\label{ad7}
  \Omega_{0}(T_0,V_0,\mu_0) = E_{0} - T_{0} S_{0} - \mu_{0} N_{0},
\end{equation}
where $S_0(T_0,V_0,\mu_0)$ and $N_0(T_0,V_0,\mu_0)$ are determined by solving the equations $T_0 = \frac{\partial E_0}{\partial S_0}$ and $\mu_0 = \frac{\partial E_0}{\partial N_0}$ at constant temperature $T_0$ and chemical potential $\mu_0$, respectively. The differential of the grand potential is:
\begin{align}\label{ad8}
   d\Omega_{0} &= -S_{0} dT_{0} - p_{0}dV_{0} - N_{0} d\mu_{0}, \\ \label{ad9}
   S_{0}  &= -\left(\frac{\partial \Omega_{0}}{\partial T_{0}}\right)_{V_{0}\mu_{0}}, \quad  p_{0} = - \left(\frac{\partial \Omega_{0}}{\partial V_{0}}\right)_{T_{0}\mu_{0}}, \nonumber \\    N_{0} &= - \left(\frac{\partial \Omega_{0}}{\partial \mu_{0}}\right)_{T_{0}V_{0}},
\end{align}
where $S_0$ is the entropy, $p_0$ is the pressure, and $N_0$ is the particle number of the system.

\paragraph{5. Microcanonical ensemble $(E_0, V_0, N_0)$}
\ The microcanonical ensemble is derived from the fundamental ensemble by replacing the variable $E_0$ with $S_0$. The thermodynamic potential for the microcanonical ensemble is the entropy $S_0$, with $E_0$ serving as the state variable:
\begin{equation}\label{ad10}
  S_{0} = S_{0}(E_0, V_0, N_0).
\end{equation}
The differential of the entropy $S_0$ with respect to the state variables $(E_0, V_0, N_0)$ is given by:
\begin{align}\label{ad11}
   dS_{0} &= \frac{1}{T_{0}} dE_{0} +\frac{p_{0}}{T_{0}} dV_{0} - \frac{\mu_{0}}{T_{0}} dN_{0}, \\ \label{ad12}
   \frac{1}{T_{0}}  &=  \left(\frac{\partial S_{0}}{\partial E_{0}}\right)_{V_{0}N_{0}}, \quad \frac{p_{0}}{T_{0}} = \left(\frac{\partial S_{0}}{\partial V_{0}}\right)_{E_{0}N_{0}}, \nonumber \\    \frac{\mu_{0}}{T_{0}} &= - \left(\frac{\partial S_{0}}{\partial N_{0}}\right)_{E_{0}V_{0}},
\end{align}
where $T_0$ is the temperature, $p_0$ is the pressure, and $\mu_0$ is the chemical potential of the system.

\subsubsection{Motion of a system in reference frame $K$ with momentum $\mathbf{P}$ as an independent state variable}
The thermodynamic system $\mathcal{A}$ is at rest in the reference frame $K_0$, which moves in a straight line at a constant velocity $\mathbf{v}$ relative to the stationary laboratory frame $K$. Here, we treat the three-momentum $\mathbf{P}$ of the system $\mathcal{A}$ as an independent state variable, making the three-velocity $\mathbf{v}$ a dependent variable. Consequently, the system is characterized by both thermodynamic state variables and the mechanical state variable $\mathbf{P}$. Below, we outline the primary ensembles of equilibrium thermodynamics for this moving system.

\paragraph{1. Fundamental ensemble $(S,V,N,\mathbf{P})$}
In the fundamental ensemble in the reference frame $K$, the variables of state of the thermodynamic system $\mathcal{A}$ are the entropy $S$, the volume $V$, the number of particles $N$ and the three-momentum $\mathbf{P}$. The fundamental thermodynamic potential is the total energy of the system:
\begin{equation}\label{ad13}
  E=E(S,V,N,\mathbf{P}).
\end{equation}
The differential of the thermodynamic potential $E$ with respect to the state variables $(S,V,N,\mathbf{P})$ is given by:
\begin{align}\label{ad14}
   dE &=  TdS - pdV + \mu dN + \mathbf{U} \cdot d\mathbf{P}, \\ \label{ad15}
   T  &\equiv  \left(\frac{\partial E}{\partial S}\right)_{VN\mathbf{P}}, \quad  p \equiv - \left(\frac{\partial E}{\partial V}\right)_{SN\mathbf{P}}, \nonumber \\    \mu &\equiv  \left(\frac{\partial E}{\partial N}\right)_{SV\mathbf{P}}, \quad   \mathbf{U} = \left(\frac{\partial E}{\partial \mathbf{P}}\right)_{SVN},
\end{align}
where $T$, $p$, $\mu$ and $\mathbf{U}$ represent the temperature, pressure, chemical potential and velocity, respectively. The dependence of the total energy (Eq. \ref{ad13}) on the momentum $\mathbf{P}$ as an independent state variable implies that the fundamental thermodynamic potential is a Hamiltonian of the system.

\paragraph{2. Canonical ensemble $(T,V,N,\mathbf{P})$}
The canonical ensemble is obtained from the fundamental ensemble by replacing the variable $S$ with $T$. The thermodynamic potential for the canonical ensemble, the Helmholtz free energy $F$, is derived from the fundamental thermodynamic potential $E$ via a Legendre transform:
\begin{equation}\label{ad16}
  F(T,V,N,\mathbf{P}) = E - T S,
\end{equation}
where $S(T,V,N,\mathbf{P})$ is determined by solving the equation $T = \frac{\partial E}{\partial S}$ at constant $T$. The differential of the Helmholtz free energy $F$ with respect to the state variables $(T,V,N,\mathbf{P})$ is given by:
\begin{align}\label{ad17}
   dF &= -S dT - pdV + \mu dN + \mathbf{U} \cdot d\mathbf{P}, \\ \label{ad18}
   S  &=  -\left(\frac{\partial F}{\partial T}\right)_{VN\mathbf{P}}, \quad  p = - \left(\frac{\partial F}{\partial V}\right)_{TN\mathbf{P}}, \nonumber \\    \mu &= \left(\frac{\partial F}{\partial N}\right)_{TV\mathbf{P}}, \quad  \mathbf{U} = \left(\frac{\partial F}{\partial \mathbf{P}}\right)_{TVN},
\end{align}
where $S$ is the entropy, $p$ is the pressure, $\mu$ is the chemical potential and $\mathbf{U}$ is the velocity of the system.

\paragraph{3. Isobaric ensemble $(T,p,N,\mathbf{P})$}
The isobaric ensemble is derived from the fundamental ensemble by replacing the variables $S$ and $V$ with $T$ and $p$, respectively. The thermodynamic potential for the isobaric ensemble, the Gibbs free energy $G$, is obtained from the fundamental thermodynamic potential $E$ through a Legendre transform:
\begin{equation}\label{ad19}
  G(T,p,N,\mathbf{P}) = E - T S + p V,
\end{equation}
where $S(T,p,N,\mathbf{P})$ and $V(T,p,N,\mathbf{P})$ are determined by solving the equations $T = \frac{\partial E}{\partial S}$ and $p = -\frac{\partial E}{\partial V}$ at constant temperature $T$ and pressure $p$, respectively. The differential of the Gibbs free energy is given by:
\begin{align}\label{ad20}
   dG &= -S dT + Vdp + \mu dN + \mathbf{U} \cdot d\mathbf{P}, \\ \label{ad21}
   S  &= -\left(\frac{\partial G}{\partial T}\right)_{pN\mathbf{P}}, \quad  V =  \left(\frac{\partial G}{\partial p}\right)_{TN\mathbf{P}}, \nonumber \\    \mu &=  \left(\frac{\partial G}{\partial N}\right)_{Tp\mathbf{P}},   \quad  \mathbf{U} = \left(\frac{\partial G}{\partial \mathbf{P}}\right)_{TpN},
\end{align}
where $S$ represents the entropy, $V$ the volume, $\mu$ the chemical potential, and $\mathbf{U}$ the velocity of the system.

\paragraph{4. Grand canonical ensemble $(T,V,\mu,\mathbf{P})$}
\ The grand canonical ensemble is obtained from the fundamental ensemble by replacing the variables $S$ and $N$ with $T$ and $\mu$, respectively. The thermodynamic potential for the grand canonical ensemble, the grand potential $\Omega$, is derived from the fundamental thermodynamic potential $E$ via a Legendre transform:
\begin{equation}\label{ad22}
  \Omega(T,V,\mu,\mathbf{P}) = E - T S - \mu N,
\end{equation}
where $S(T,V,\mu,\mathbf{P})$ and $N(T,V,\mu,\mathbf{P})$ are determined by solving the equations $T = \frac{\partial E}{\partial S}$ and $\mu = \frac{\partial E}{\partial N}$ at constant temperature $T$ and chemical potential $\mu$, respectively. The differential of the grand potential is:
\begin{align}\label{ad23}
   d\Omega &= -S dT - pdV - N d\mu + \mathbf{U} \cdot d\mathbf{P}, \\ \label{ad24}
   S  &= -\left(\frac{\partial \Omega}{\partial T}\right)_{V\mu\mathbf{P}}, \quad  p = - \left(\frac{\partial \Omega}{\partial V}\right)_{T\mu\mathbf{P}}, \nonumber \\    N &= - \left(\frac{\partial \Omega}{\partial \mu}\right)_{TV\mathbf{P}},  \quad  \mathbf{U} = \left(\frac{\partial \Omega}{\partial \mathbf{P}}\right)_{TV\mu},
\end{align}
where $S$ is the entropy, $p$ is the pressure, $N$ is the particle number, and $\mathbf{U}$ the velocity of the system.

\paragraph{5. Microcanonical ensemble $(E,V,N,\mathbf{P})$}
\ The microcanonical ensemble is obtained from the fundamental ensemble by replacing the variable $E$ with $S$. The thermodynamic potential for the microcanonical ensemble is the entropy $S$, with $E$ serving as the state variable:
\begin{equation}\label{ad25}
  S = S(E,V,N,\mathbf{P}).
\end{equation}
The differential of the entropy $S$ with respect to the state variables $(E,V,N,\mathbf{P})$ is given by:
\begin{align}\label{ad26}
   dS &= \frac{1}{T} dE +\frac{p}{T} dV - \frac{\mu}{T} dN - \frac{1}{T} \ \mathbf{U} \cdot d\mathbf{P}, \\ \label{ad27}
   \frac{1}{T}  &=  \left(\frac{\partial S}{\partial E}\right)_{VN\mathbf{P}}, \quad \frac{p}{T} = \left(\frac{\partial S}{\partial V}\right)_{EN\mathbf{P}}, \nonumber \\    \frac{\mu}{T} &= - \left(\frac{\partial S}{\partial N}\right)_{EV\mathbf{P}}, \quad \frac{\mathbf{U}}{T} = - \left(\frac{\partial S}{\partial \mathbf{P}}\right)_{EVN},
\end{align}
where $T$ is the temperature, $p$ is the pressure, $\mu$ is the chemical potential, and $\mathbf{U}$ is the velocity of the system.

\subsubsection{Motion of a system in reference frame $K$ with velocity $\mathbf{v}$ as an independent state variable}
Here, we treat the three-velocity $\mathbf{v}$ of the system $\mathcal{A}$ as an independent state variable, with the three-momentum $\mathbf{P}$ as a dependent variable. Below, we describe the primary ensembles of equilibrium thermodynamics for this moving system.

\paragraph{1. Conjugate fundamental ensemble $(S,V,N,\mathbf{v})$}
In the conjugate fundamental ensemble in the reference frame $K$, the variables of state of the thermodynamic system $\mathcal{A}$ are the entropy $S$, the volume $V$, the number of particles $N$ and the three-velocity $\mathbf{v}$. We define the conjugate fundamental thermodynamic potential as the negative of the system's Lagrangian (cf. Eq.~\eqref{9}):
\begin{equation}\label{ad28}
  \bar{L}(S,V,N,\mathbf{v})=E-\mathbf{v} \cdot \mathbf{P}  = \frac{u_{\mu}P^{\mu}}{\gamma},
\end{equation}
where $L=-\bar{L}$ represents the Lagrangian of the thermodynamic system~\cite{Landau}. The differential of the thermodynamic potential $\bar{L}$ with respect to the state variables $(S,V,N,\mathbf{v})$ is given by:
\begin{align}\label{ad29}
   d\bar{L} &=  TdS - pdV + \mu dN - \mathbf{G} \cdot d\mathbf{v}, \\ \label{ad30}
   T  &\equiv  \left(\frac{\partial \bar{L}}{\partial S}\right)_{VN\mathbf{v}}, \quad  p \equiv - \left(\frac{\partial \bar{L}}{\partial V}\right)_{SN\mathbf{v}}, \nonumber \\    \mu &\equiv  \left(\frac{\partial \bar{L}}{\partial N}\right)_{SV\mathbf{v}}, \quad   \mathbf{G} = -\left(\frac{\partial \bar{L}}{\partial \mathbf{v}}\right)_{SVN},
\end{align}
where $T$, $p$, $\mu$ and $\mathbf{G}$ represent the temperature, pressure, chemical potential and momentum, respectively.

\paragraph{2. Conjugate canonical ensemble $(T,V,N,\mathbf{v})$}
The conjugate canonical ensemble is obtained from the conjugate fundamental ensemble by replacing the variable $S$ with $T$. The thermodynamic potential for the conjugate canonical ensemble, the conjugate free energy $\bar{F}$, is derived from the conjugate fundamental thermodynamic potential $\bar{L}$ via a Legendre transformation:
\begin{equation}\label{ad31}
  \bar{F}(T,V,N,\mathbf{v}) = \bar{L} - T S = \frac{u_{\mu}P^{\mu}}{\gamma} -TS,
\end{equation}
where $S(T,V,N,\mathbf{v})$ is determined by solving the equation $T = \frac{\partial \bar{L}}{\partial S}$ at constant $T$. The differential of the conjugate free energy $\bar{F}$ with respect to the state variables $(T,V,N,\mathbf{v})$ is given by:
\begin{align}\label{ad32}
   d\bar{F} &= -S dT - pdV + \mu dN - \mathbf{G} \cdot d\mathbf{v}, \\ \label{ad33}
   S  &=  -\left(\frac{\partial \bar{F}}{\partial T}\right)_{VN\mathbf{v}}, \quad  p = - \left(\frac{\partial \bar{F}}{\partial V}\right)_{TN\mathbf{v}}, \nonumber \\    \mu &= \left(\frac{\partial \bar{F}}{\partial N}\right)_{TV\mathbf{v}}, \quad  \mathbf{G} = -\left(\frac{\partial \bar{F}}{\partial \mathbf{v}}\right)_{TVN},
\end{align}
where $S$ is the entropy, $p$ is the pressure, $\mu$ is the chemical potential and $\mathbf{G}$ is the momentum of the system.

\paragraph{3. Conjugate isobaric ensemble $(T,p,N,\mathbf{v})$}
The conjugate isobaric ensemble is derived from the conjugate fundamental ensemble by replacing the variables $S$ and $V$ with $T$ and $p$, respectively. The thermodynamic potential for the conjugate isobaric ensemble, the conjugate free energy $\bar{G}$, is obtained from the conjugate fundamental thermodynamic potential $\bar{L}$ through a Legendre transformation:
\begin{equation}\label{ad34}
  \bar{G}(T,p,N,\mathbf{v}) = \bar{L} - T S + p V = \frac{u_{\mu}P^{\mu}}{\gamma} -TS +pV,
\end{equation}
where $S(T,p,N,\mathbf{v})$ and $V(T,p,N,\mathbf{v})$ are determined by solving the equations $T = \frac{\partial \bar{L}}{\partial S}$ and $p = -\frac{\partial \bar{L}}{\partial V}$ at constant temperature $T$ and pressure $p$, respectively. The differential of the conjugate free energy is given by:
\begin{align}\label{ad35}
   d\bar{G} &= -S dT + Vdp + \mu dN - \mathbf{G} \cdot d\mathbf{v}, \\ \label{ad36}
   S  &= -\left(\frac{\partial \bar{G}}{\partial T}\right)_{pN\mathbf{v}}, \quad  V =  \left(\frac{\partial \bar{G}}{\partial p}\right)_{TN\mathbf{v}}, \nonumber \\    \mu &=  \left(\frac{\partial \bar{G}}{\partial N}\right)_{Tp\mathbf{v}},   \quad  \mathbf{G} = -\left(\frac{\partial \bar{G}}{\partial \mathbf{v}}\right)_{TpN},
\end{align}
where $S$ represents the entropy, $V$ the volume, $\mu$ the chemical potential, and $\mathbf{G}$ the momentum of the system.

\paragraph{4. Conjugate grand canonical ensemble $(T,V,\mu,\mathbf{v})$}
The conjugate grand canonical ensemble is obtained from the conjugate fundamental ensemble by replacing the variables $S$ and $N$ with $T$ and $\mu$, respectively. The thermodynamic potential for the conjugate grand canonical ensemble, the conjugate grand potential $\bar{\Omega}$, is derived from the conjugate fundamental thermodynamic potential $\bar{L}$ via a Legendre transformation:
\begin{equation}\label{ad37}
  \bar{\Omega}(T,V,\mu,\mathbf{v}) = \bar{L} - T S - \mu N = \frac{u_{\mu}P^{\mu}}{\gamma}- T S - \mu N,
\end{equation}
where $S(T,V,\mu,\mathbf{v})$ and $N(T,V,\mu,\mathbf{v})$ are determined by solving the equations $T = \frac{\partial \bar{L}}{\partial S}$ and $\mu = \frac{\partial \bar{L}}{\partial N}$ at constant temperature $T$ and chemical potential $\mu$, respectively. The differential of the conjugate grand potential is:
\begin{align}\label{ad38}
   d\bar{\Omega} &= -S dT - pdV - N d\mu - \mathbf{G} \cdot d\mathbf{v}, \\ \label{ad39}
   S  &= -\left(\frac{\partial \bar{\Omega}}{\partial T}\right)_{V\mu\mathbf{v}}, \quad  p = - \left(\frac{\partial \bar{\Omega}}{\partial V}\right)_{T\mu\mathbf{v}}, \nonumber \\    N &= - \left(\frac{\partial \bar{\Omega}}{\partial \mu}\right)_{TV\mathbf{v}},  \quad  \mathbf{G} = -\left(\frac{\partial \bar{\Omega}}{\partial \mathbf{v}}\right)_{TV\mu},
\end{align}
where $S$ is the entropy, $p$ is the pressure, $N$ is the particle number, and $\mathbf{G}$ the momentum of the system.

\paragraph{5. Conjugate microcanonical ensemble $(\bar{L},V,N,\mathbf{v})$}
The conjugate microcanonical ensemble is obtained from the conjugate fundamental ensemble by replacing the variable $\bar{L}$ with $S$. The thermodynamic potential for the conjugate microcanonical ensemble is the entropy $S$, with $\bar{L}$ serving as the state variable:
\begin{equation}\label{ad40}
  S = S(\bar{L},V,N,\mathbf{v}).
\end{equation}
The differential of the entropy $S$ with respect to the state variables $(\bar{L},V,N,\mathbf{v})$ is given by:
\begin{align}\label{ad41}
   dS &= \frac{1}{T} d\bar{L} +\frac{p}{T} dV - \frac{\mu}{T} dN + \frac{1}{T} \ \mathbf{G} \cdot d\mathbf{v}, \\ \label{ad42}
   \frac{1}{T}  &=  \left(\frac{\partial S}{\partial \bar{L}}\right)_{VN\mathbf{v}}, \quad \frac{p}{T} = \left(\frac{\partial S}{\partial V}\right)_{\bar{L}N\mathbf{v}}, \nonumber \\    \frac{\mu}{T} &= - \left(\frac{\partial S}{\partial N}\right)_{\bar{L}V\mathbf{v}}, \quad \frac{\mathbf{G}}{T} =  \left(\frac{\partial S}{\partial \mathbf{v}}\right)_{\bar{L}VN},
\end{align}
where $T$ is the temperature, $p$ is the pressure, $\mu$ is the chemical potential, and $\mathbf{G}$ is the momentum of the system.

In equilibrium thermodynamics within the reference frame $K$, a thermodynamic system $\mathcal{A}$ is fully described by its fundamental thermodynamic potential $E$, the system's total energy, expressed as a function of state variables $(S, V, N, \mathbf{P})$. The system's total energy $E$, expressed as a function of momentum $\mathbf{P}$ as an independent state variable, corresponds to the system's Hamiltonian. The fundamental ensemble can be transformed into the canonical, isobaric, and grand canonical ensembles through Legendre transformations. These transformations convert the energy function $E(S, V, N, \mathbf{P})$ into the Helmholtz free energy $F(T, V, N, \mathbf{P})$, Gibbs free energy $G(T, p, N, \mathbf{P})$, and grand potential $\Omega(T, V, \mu, \mathbf{P})$, respectively, by substituting the appropriate state variables. The microcanonical ensemble is obtained by replacing $E$ with the entropy $S$.

In this paper, we define the conjugate fundamental thermodynamic potential $\bar{L}(S, V, N, \mathbf{v})$, where the velocity $\mathbf{v}$ is an independent state variable, as the negative of the system's Lagrangian. From this conjugate fundamental ensemble, we derive the conjugate canonical, isobaric, and grand canonical ensembles by substituting the appropriate state variables and transforming the conjugate thermodynamic potential $\bar{L}(S, V, N, \mathbf{v})$ into the conjugate free energies $\bar{F}(T, V, N, \mathbf{v})$, $\bar{G}(T, p, N, \mathbf{v})$, and the conjugate grand potential $\bar{\Omega}(T, V, \mu, \mathbf{v})$ via Legendre transformations. The conjugate microcanonical ensemble is obtained by replacing the state variable $\bar{L}$ with $S$.

In this study, we derive relativistic transformations for thermodynamic quantities within the fundamental ensemble and its conjugate ensemble. We introduce the concepts of the conjugate ensemble and conjugate thermodynamic potential, specifically when velocity is an independent state variable, to distinguish them from the ensembles and thermodynamic potentials associated with momentum as an independent state variable.

Our research focuses solely on thermodynamic systems at the nuclear and hadronic scales. In high-energy physics, Lorentz transformations are effectively used to describe non-point relativistic systems, such as nuclei (hadron systems) or hadrons (parton systems). Despite their non-point nature, these transformations are widely applied, as extensively discussed in textbooks on relativistic nuclear physics and the literature on heavy-ion and proton-proton collisions.

In this study, the infinitesimal change in velocity (or momentum) is considered according to the rules of equilibrium thermodynamics, expressed as $ d\mathbf{v} = \mathbf{v}_2 - \mathbf{v}_1 $, where $ \mathbf{v}_1 $ represents the velocity in the initial state and $\mathbf{v}_2 $ denotes the velocity in the final state. The transition between these states is achieved by first constructing a thermodynamic system in the initial state, then configuring it in the final state, and subsequently replacing them~\cite{Kvasnikov}.

\section{Relativistic transformations of thermodynamic quantities}
%\subsection{Fundamental ensemble $(S,V,N,\mathbf{P})$}\label{sec3}
\subsection{Momentum $\mathbf{P}$ as an independent state variable: Non-Planck transformations}\label{sec3}
Consider a thermodynamic system $\mathcal{A}$ in \textit{the fundamental ensemble} $(S, V, N, \mathbf{P})$ within the reference frame $K$. In the reference frame $K_0$, which moves relative to the laboratory frame $K$, the system is at rest. The three-momentum $\mathbf{P}$ serves as an independent state variable. Thus, the fundamental thermodynamic potential $E$ (total energy) in the reference frame $K$ is the Hamiltonian function. The Hamiltonian mechanics requires that the Hamiltonian should be a function of momentum as an independent state variable.

In the reference frame $K_0$, the fundamental thermodynamic potential of the thermodynamic system $\mathcal{A}$ is represented by its rest energy $E_0$ (internal energy), as given in Eq.~(\ref{9a}). Using Eqs.~(\ref{5})--(\ref{8}), we derive the following expression for the Lorentz factor:
\begin{equation}\label{11}
  \gamma = \sqrt{1+\frac{\mathbf{P}^{2}}{E_{0}^{2}}}.
\end{equation}
When transitioning from the reference frame $K$ to the reference frame $K_0$, the independent state variables transform as follows:
\begin{equation}\label{10a}
  S_{0} = S, \quad  N_{0} = N, \quad V_{0} = \gamma V.
\end{equation}
The entropy $S$ and the number of particles $N$ of the thermodynamic system $\mathcal{A}$ are relativistic invariants~\cite{Haar,Farias}. However, the volume $V$ of the system in the reference frame $K$ undergoes Lorentz contraction~\cite{Haar,Farias,Landau}. Then, the internal energy (\ref{9a}) is transformed as
\begin{equation}\label{ad43}
  E_{0}(S_{0},V_{0},N_{0}) = E_{0}(S,\gamma V,N).
\end{equation}
Substituting Eq.~(\ref{ad43}) into Eq.~(\ref{11}), we obtain the following equation:
\begin{equation}\label{ad44}
  \gamma = \left[1+\frac{\mathbf{P}^{2}}{E_{0}^{2}(S,\gamma V,N)}\right]^{1/2}.
\end{equation}
This is an equation of self-consistency. Solving Eq.~(\ref{ad44}) with respect to the unknown $\gamma$ at fixed values of variables $(S,V,N,\mathbf{P})$, we obtain the function:
\begin{equation}\label{ad45}
  \gamma=\gamma(S,V,N,\mathbf{P}).
\end{equation}
Then, the total energy (\ref{6}) (fundamental thermodynamic potential) of the thermodynamic system $\mathcal{A}$ in the reference frame $K$ can be written as:
\begin{equation}\label{23a}
  E = \gamma  E_{0}(S_{0},V_{0},N_{0})  = \sqrt{\mathbf{P}^{2}+E_{0}^{2}(S_{0},V_{0},N_{0})},
\end{equation}
where $S_{0}$, $V_{0}$ and $N_{0}$ are defined in Eq.~(\ref{10a}). Differentiating $E$ with respect to the independent state variables $(S,V,N,\mathbf{P})$ of the thermodynamic system $\mathcal{A}$ in the reference frame $K$, and applying Eqs.~(\ref{5}), (\ref{8}), (\ref{9a}), (\ref{2}), (\ref{11}), (\ref{10a}), and (\ref{ad44}), we obtain
\begin{align}\label{a5z}
  dE &=   \frac{\alpha T_{0}}{\gamma} dS -  \alpha p_{0} dV + \frac{\alpha \mu_{0}}{\gamma} dN \nonumber \\ &+ \alpha (1-\frac{p_{0}V_{0}}{E_{0}}) \ \mathbf{v} \cdot d\mathbf{P},
\end{align}
where
\begin{equation}\label{23}
  \alpha \equiv \frac{1}{1-\frac{V_{0}p_{0}}{E_{0}}\frac{\mathbf{P}^{2}}{E^{2}}} = \frac{1}{1-\frac{V_{0}p_{0}}{E_{0}}\mathbf{v}^{2}}.
\end{equation}
The fundamental equation of thermodynamics in reference frame $K$ is expressed as:
\begin{equation}\label{a5}
  dE =  T dS -  p dV +  \mu dN +  \mathbf{U} \cdot d\mathbf{P},
\end{equation}
where the thermodynamic quantities are defined as (cf. Eq.~(\ref{ad15})):
\begin{align}\label{20}
 T \equiv &  \left(\frac{\partial E}{\partial S}\right)_{VN\mathbf{P}}, \quad  p \equiv   - \left(\frac{\partial E}{\partial V}\right)_{SN\mathbf{P}}, \nonumber \\
 \mu \equiv & \left(\frac{\partial E}{\partial N}\right)_{SV\mathbf{P}}, \quad  \mathbf{U} \equiv  \left(\frac{\partial E}{\partial \mathbf{P}}\right)_{SVN}.
\end{align}
Here, $T$, $p$, $\mu$ and $\mathbf{U}$ represent the temperature, pressure, chemical potential, and generalized velocity of the thermodynamic system $\mathcal{A}$ in the reference frame $K$. By comparing Eqs.~(\ref{a5}) and (\ref{a5z}), we derive the relativistic transformations for the thermodynamic quantities defined in Eqs.~(\ref{20}) and (\ref{2}):
\begin{align}\label{a7}
  T &=  \frac{\alpha T_{0}}{\gamma}, \qquad p=\alpha p_{0}, \qquad \mu =\frac{\alpha \mu_{0}}{\gamma}, \nonumber \\
  \mathbf{U} &= \alpha \left(1-\frac{p_{0}V_{0}}{E_{0}}\right)\mathbf{v},
\end{align}
where
\begin{equation}\label{v2}
  \mathbf{v} =  \frac{\mathbf{P}}{E(S,V,N,\mathbf{P})}.
\end{equation}
These transformations of the thermodynamic quantities differ from the Planck transformations by a factor of $\alpha$.

It can be readily shown that, in the state variables $(S, V, N, \mathbf{P})$, the following equation holds (cf. Eq.~\eqref{3a}):
\begin{equation}\label{n1}
  dE_{0} = \gamma (dE-\mathbf{v} \cdot d\mathbf{P}) = u_{\mu} dP^{\mu},
\end{equation}
where $dP^{\mu}=(dE,d\mathbf{P})$. The rest volume $V_{0}$ at fixed values of $(S,V,N,\mathbf{P})$ takes the form:
\begin{equation}\label{a1a}
  V_{0}=\gamma(S,V,N,\mathbf{P}) V.
\end{equation}
For the differential $dV_{0}$ in the variables $(S, V, N, \mathbf{P})$, the following relation is satisfied:
\begin{align}\label{15ba}
   dV_{0} &= -\alpha \frac{V_{0} T_{0}}{E_{0}} \frac{\mathbf{P}^{2}}{E^{2}} dS + \alpha \gamma dV
   - \alpha \frac{V_{0} \mu_{0}}{E_{0}}  \frac{\mathbf{P}^{2}}{E^{2}} dN \nonumber \\ &+ \alpha \frac{V_{0}}{E} \ \mathbf{v} \cdot d\mathbf{P}.
\end{align}

The dependence of the $\gamma$-factor and velocity $\mathbf{v}$ on the thermodynamic state variables at fixed momentum $\mathbf{P}$ lacks experimental confirmation. Therefore, in this work, the derivation of thermodynamic quantities and relativistic transformations from the system's total energy $E$ at fixed momentum $\mathbf{P}$ is performed purely mathematically for completeness, without consideration of physical implications.

\paragraph{Transition to state variables $(S, V, N, \mathbf{v})$}
To treat velocity $\mathbf{v}$ as an independent state variable, the momentum $\mathbf{P}$ in the thermodynamic potential $E(S, V, N, \mathbf{P})$ must be replaced with velocity $\mathbf{v}$. This transformation, rooted in thermodynamic principles, involves converting $E(S, V, N, \mathbf{P})$ into a new potential, $\tilde{L}(S, V, N, \mathbf{v})$, using a Legendre transformation:
\begin{equation}\label{ad1x}
  \tilde{L}(S,V,N,\mathbf{v}) = E - \frac{\partial E}{\partial \mathbf{P}} \cdot \mathbf{P} = E - \mathbf{U} \cdot \mathbf{P}.
\end{equation}
By substituting Eq.~(\ref{a7}) into Eq.~(\ref{ad1x}) and applying Eqs.~(\ref{6}) and (\ref{23}), we derive:
\begin{equation}\label{ad2x}
  \tilde{L} = \frac{\alpha E_{0}}{\gamma} = \alpha \ \frac{u_{\mu}P^{\mu}}{\gamma}.
\end{equation}
This expression represents the thermodynamic potential $\tilde{L}$ in terms of the state variables $(S, V, N, \mathbf{v})$, obtained from the fundamental thermodynamic potential $E(S, V, N, \mathbf{P})$ (the Hamiltonian function) through Legendre transformations. In the state variables $(S, V, N, \mathbf{v})$, the total energy $E(S, V, N, \mathbf{v})$, derived from the fundamental thermodynamic potential $E(S, V, N, \mathbf{P})$ by substituting the variable $\mathbf{P}$ with $\mathbf{v}$, does not constitute a thermodynamic potential. As discussed in Subsection~\ref{sec5}, this function gives rise to the Ott transformations.

\subsection{Velocity $\mathbf{v}$ as an independent state variable: Planck transformations}\label{sec4}
\subsubsection{Conjugate fundamental ensemble $(S,V,N,\mathbf{v})$}
Consider a thermodynamic system $\mathcal{A}$ in the conjugate fundamental ensemble $(S, V, N, \mathbf{v})$ within the reference frame $K$. The system is at rest in the reference frame $K_0$, which moves relative to the laboratory frame $K$. The velocity $\mathbf{v}$ is an independent state variable. In this study, we define the conjugate fundamental thermodynamic potential $\bar{L}$ as the negative Lagrangian function. In Lagrangian mechanics, the Lagrangian must be a function of velocity as an independent state variable.

In the reference frame $K_0$, the rest energy $E_0$, as defined in Eq.~(\ref{9a}), serves as the fundamental thermodynamic potential for the thermodynamic system $\mathcal{A}$. Since the velocity $\mathbf{v}$ is an independent state variable, the Lorentz factor, given by Eq.~(\ref{5}), depends solely on velocity:
\begin{equation}\label{ad46}
  \gamma(\mathbf{v}) = \frac{1}{\sqrt{1-\mathbf{v}^{2}}}.
\end{equation}
The thermodynamic state variables of the conjugate fundamental ensemble and the rest energy are transformed according to Eqs.~(\ref{10a}) and (\ref{ad43}). The thermodynamic potential of the conjugate fundamental ensemble is defined by the negative Lagrangian. Substituting Eqs.~(\ref{6}), (\ref{9a}) into Eq.~(\ref{ad28}), we obtain~\cite{Agmon}
\begin{equation}\label{44a}
  \bar{L} = E - \mathbf{v} \cdot \mathbf{P}= \frac{u_{\mu}P^{\mu}}{\gamma(\mathbf{v})} = \frac{E_{0}(S_{0},V_{0},N_{0})}{\gamma(\mathbf{v})},
\end{equation}
where $S_{0}$, $V_{0}$ and $N_{0}$ are defined in Eq.~(\ref{10a}). The rest volume $V_{0}$ at fixed values of $(V,\mathbf{v})$ takes the form:
\begin{equation}\label{b1}
  V_{0}=\gamma(\mathbf{v}) V.
\end{equation}
Differentiating the volume expression (\ref{b1}) with respect to the variables $(V,\mathbf{v})$, and using the independence condition $\partial V/\partial \mathbf{v} =0$, we obtain:
\begin{equation}\label{39}
  dV_{0} = \gamma dV + \gamma \frac{V_{0}}{E_{0}} \ \mathbf{P} \cdot d\mathbf{v}.
\end{equation}
Differentiating $\bar{L}$ with respect to the independent state variables $(S,V,N,\mathbf{v})$ of the thermodynamic system $\mathcal{A}$ in the reference frame $K$, and applying Eqs.~(\ref{6}), (\ref{9a}), (\ref{2}), (\ref{10a}), (\ref{ad46}), and (\ref{b1}), we obtain:
\begin{equation}\label{41z}
  d\bar{L} = \frac{T_{0}}{\gamma} dS - p_{0} dV + \frac{\mu_{0}}{\gamma} dN - (1+\frac{p_{0}V_{0}}{E_{0}}) \mathbf{P} \cdot d\mathbf{v}.
\end{equation}
The fundamental equation of thermodynamics in reference frame $K$, expressed in terms of the variable $\mathbf{v}$, is given by:
\begin{equation}\label{41}
  d\bar{L} = T dS - p dV + \mu dN - \mathbf{G} \cdot d\mathbf{v},
\end{equation}
where the thermodynamic quantities are defined as (cf. Eq.~(\ref{ad30}))
\begin{align}\label{42}
 T  &\equiv  \left(\frac{\partial \bar{L}}{\partial S}\right)_{VN\mathbf{v}}, \quad  p \equiv - \left(\frac{\partial \bar{L}}{\partial V}\right)_{SN\mathbf{v}}, \nonumber \\    \mu &\equiv  \left(\frac{\partial \bar{L}}{\partial N}\right)_{SV\mathbf{v}}, \quad   \mathbf{G} = -\left(\frac{\partial \bar{L}}{\partial \mathbf{v}}\right)_{SVN}.
\end{align}
Here, $T$, $p$, $\mu$ and $\mathbf{G}$ represent the temperature, pressure, chemical potential, and generalized momentum, respectively, of the thermodynamic system $\mathcal{A}$ in the reference frame $K$. By comparing Eqs.~(\ref{41}) and (\ref{41z}), we derive the Planck transformations~\cite{Haar,Farias,Mosengeil,Planck,Einstein,Agmon,Parvan2019} for the thermodynamic quantities defined in Eqs.~(\ref{42}) and (\ref{2}):
\begin{align}\label{39b}
  T  &=  \frac{T_{0}}{\gamma}, \qquad    p =  p_{0}, \qquad
  \mu = \frac{\mu_{0}}{\gamma}, \nonumber \\   \mathbf{G} &= \left(1+\frac{p_{0}V_{0}}{E_{0}}\right) \mathbf{P},
\end{align}
where
\begin{equation}\label{44b}
  \mathbf{P} = \mathbf{v} E, \quad E = \gamma  E_{0}(S_{0},V_{0},N_{0}).
\end{equation}

The Planck transformation corresponds to the conjugate fundamental thermodynamic potential $\bar{L}$, defined in terms of the state variables $(S, V, N, \mathbf{v})$ as the negative of the Lagrangian function. See~\ref{App}. This potential is distinct from the thermodynamic potential $\tilde{L}(S, V, N, \mathbf{v})$, which is derived from fundamental thermodynamic potential $E(S, V, N, \mathbf{P})$ (the Hamiltonian function) through Legendre transformations in the same state variables $(S, V, N, \mathbf{v})$. The distinction between these potentials arises from the factor $\alpha$ (see Eqs.~(\ref{ad2x}) and (\ref{44a})):
\begin{equation}\label{ad6x}
  \tilde{L}=\alpha \bar{L}.
\end{equation}
It can be readily demonstrated that, for the state variables $(S, V, N, \mathbf{v})$, the following equation holds:
\begin{equation}\label{n2}
  dE_{0} = \gamma (d\bar{L}+\mathbf{P} \cdot d\mathbf{v}).
\end{equation}

\begin{table*}
\caption{Overview of the formulas for the thermodynamic potentials and their associated thermodynamic quantities, expressed in terms of the state variables $(S, V, N, \mathbf{P})$ and $(S, V, N, \mathbf{v})$. Here, $S_0 = S$, $V_0 = \gamma V$, and $N_0 = N$.}\label{tab:1}
\begin{tabular}{|p{5.3cm}|p{5.3cm}|p{5.3cm}|}
\hline
 Non-Planck transformations  &     Planck transformations             &     Ott transformations          \\
\hline
$\mathbf{P}=const$        & $\mathbf{v}=const$           & $\mathbf{v}=const$  \\
\hline
 $(S,V,N,\mathbf{P})$     & $(S,V,N,\mathbf{v})$         & $(S,V,N,\mathbf{v})$ \\
\hline
 $\gamma=\gamma(S,V,N,\mathbf{P})$ & $\gamma(\mathbf{v}) = \frac{1}{\sqrt{1-\mathbf{v}^{2}}}$  &  $\gamma(\mathbf{v}) = \frac{1}{\sqrt{1-\mathbf{v}^{2}}}$  \\
\hline
$E = \sqrt{\mathbf{P}^{2}+E_{0}^{2}(S_{0},V_{0},N_{0})} $ & $ \bar{L}=\frac{E_{0}(S_{0},V_{0},N_{0})}{\gamma(\mathbf{v})}$ &$E = \gamma(\mathbf{v}) E_{0}(S_{0},V_{0},N_{0})$  \\
\hline
$T= \left(\frac{\partial E}{\partial S}\right)_{VN\mathbf{P}} = \frac{\alpha T_{0}}{\gamma}$ & $T = \left(\frac{\partial \bar{L}}{\partial S}\right)_{VN\mathbf{v}} = \frac{T_{0}}{\gamma}$ & $ T_{*}= \left(\frac{\partial E}{\partial S}\right)_{VN\mathbf{v}} =  \gamma T_{0}$  \\

$p = -\left(\frac{\partial E}{\partial V}\right)_{SN\mathbf{P}}= \alpha p_{0}$  &  $p = -\left(\frac{\partial \bar{L}}{\partial V}\right)_{SN\mathbf{v}}=p_{0}$ &  $p_{*} = -\left(\frac{\partial E}{\partial V}\right)_{SN\mathbf{v}} = \gamma^{2} p_{0}$   \\

$\mu= \left(\frac{\partial E}{\partial N}\right)_{SV\mathbf{P}} = \frac{\alpha\mu_{0}}{\gamma}$ & $\mu = \left(\frac{\partial \bar{L}}{\partial N}\right)_{SV\mathbf{v}} = \frac{\mu_{0}}{\gamma}$ & $ \mu_{*}= \left(\frac{\partial E}{\partial N}\right)_{SV\mathbf{v}} =  \gamma \mu_{0}$  \\
\hline
\end{tabular}
\end{table*}

\paragraph{Transition to state variables $(S, V, N, \mathbf{P})$}
To treat momentum $\mathbf{P}$ as an independent state variable, the velocity $\mathbf{v}$ in the conjugate fundamental thermodynamic potential $\bar{L}(S, V, N, \mathbf{v})$ must be replaced with momentum $\mathbf{P}$. This transformation involves converting $\bar{L}(S, V, N, \mathbf{v})$ into a new potential, $\bar{E}(S, V, N, \mathbf{P})$, using a Legendre transformation:
\begin{equation}\label{ad3x}
  \bar{E}(S, V, N, \mathbf{P}) = \bar{L} - \frac{\partial \bar{L}}{\partial \mathbf{v}} \cdot \mathbf{v} = \bar{L} + \mathbf{G} \cdot \mathbf{v}.
\end{equation}
By substituting Eq.~(\ref{39b}) into Eq.~(\ref{ad3x}) and applying Eqs.~(\ref{6}), (\ref{ad46}) and (\ref{44a}), we derive:
\begin{equation}\label{ad4x}
  \bar{E} = \gamma E_{0} \left(1+\frac{p_{0}V_{0}}{E_{0}}\mathbf{v}^{2}\right).
\end{equation}
This equation corresponds exactly to Eq.~(69.11) in Tolman~\cite{Tolman} (see also Refs.~\cite{Callen,Agmon}). The potential $\bar{E}$, obtained through Legendre transformations from the conjugate fundamental potential $\bar{L}(S, V, N, \mathbf{v})$ (the negative Lagrangian) in the state variables $(S, V, N, \mathbf{P})$, is distinct from the fundamental thermodynamic potential $E(S, V, N, \mathbf{P})$ (the Hamiltonian) in the same state variables $(S, V, N, \mathbf{P})$ (see Eqs.~(\ref{ad4x}) and (\ref{23a})).

\subsubsection{Conjugate canonical ensemble $(T,V,N,\mathbf{v})$}
In the conjugate canonical ensemble $(T, V, N, \mathbf{v})$, the thermodynamic potential $\bar{F}$ and the quantities $S$, $p$, and $\mu$ are determined by Eqs.~(\ref{ad31}) and (\ref{ad33}), respectively. By substituting Eq.~(\ref{44a}) and the relations $S = S_0$ and $T = T_0/\gamma$ into Eq.~(\ref{ad31}), we obtain:
\begin{equation}\label{1ad}
  \bar{F} = \bar{L} -TS = \frac{F_{0}}{\gamma},
\end{equation}
where $F_0 = E_0 - T_0 S_0$ (see Eq.~(\ref{ad1})). Further, substituting Eq.~(\ref{1ad}) and the relations $T = T_0/\gamma$, $V = V_0/\gamma$, and $N = N_0$ into Eq.~(\ref{ad33}), and using Eq.~(\ref{ad3}), we derive:
\begin{equation}\label{2ad}
  S=S_{0}, \qquad p=p_{0}, \qquad \mu=\frac{\mu_{0}}{\gamma}.
\end{equation}
Thus, the Planck transformations for temperature, pressure, and chemical potential hold in the conjugate canonical ensemble $(T, V, N, \mathbf{v})$.

\subsubsection{Conjugate isobaric ensemble $(T,p,N,\mathbf{v})$}
In the conjugate isobaric ensemble $(T, p, N, \mathbf{v})$, the thermodynamic potential $\bar{G}$ and the quantities $S$, $V$, and $\mu$ are determined using Eqs.~(\ref{ad34}) and (\ref{ad36}), respectively. By substituting Eq.~(\ref{44a}) along with the relations $S = S_0$, $T = T_0/\gamma$, $p = p_0$, and $V = V_0/\gamma$ into Eq.~(\ref{ad34}), we obtain:
\begin{equation}\label{3ad}
  \bar{G} = \bar{L} -TS + pV = \frac{G_{0}}{\gamma},
\end{equation}
where $G_0 = E_0 - T_0 S_0 + p_0 V_0$ (see Eq.~(\ref{ad4})). Furthermore, by substituting Eq.~(\ref{3ad}) and the relations $T = T_0/\gamma$, $p = p_0$, and $N = N_0$ into Eq.~(\ref{ad36}), and applying Eq.~(\ref{ad6}), we derive:
\begin{equation}\label{4ad}
  S=S_{0}, \qquad V=\frac{V_{0}}{\gamma}, \qquad \mu=\frac{\mu_{0}}{\gamma}.
\end{equation}
Thus, the Planck transformations for temperature, pressure, and chemical potential are valid in the conjugate isobaric ensemble $(T, p, N, \mathbf{v})$.

\subsubsection{Conjugate grand canonical ensemble $(T,V,\mu,\mathbf{v})$}
In the conjugate grand canonical ensemble $(T,V,\mu,\mathbf{v})$, the thermodynamic potential $\bar{\Omega}$ and the quantities $S$, $p$, and $N$ are determined using Eqs.~(\ref{ad37}) and (\ref{ad39}), respectively. By substituting Eq.~(\ref{44a}) along with the relations $S = S_0$, $T = T_0/\gamma$, $\mu = \mu_0/\gamma$, and $N = N_0$ into Eq.~(\ref{ad37}), we obtain:
\begin{equation}\label{5ad}
  \bar{\Omega} = \bar{L} -TS - \mu N = \frac{\Omega_{0}}{\gamma},
\end{equation}
where $\Omega_0 = E_0 - T_0 S_0 - \mu_0 N_0$ (see Eq.~(\ref{ad7})). Furthermore, by substituting Eq.~(\ref{5ad}) and the relations $T = T_0/\gamma$, $V = V_0/\gamma$, and $\mu = \mu_0/\gamma$ into Eq.~(\ref{ad39}), and applying Eq.~(\ref{ad9}), we derive:
\begin{equation}\label{6ad}
  S=S_{0}, \qquad p = p_0, \qquad N = N_0.
\end{equation}
Thus, the Planck transformations for temperature, pressure, and chemical potential are valid in the conjugate grand canonical ensemble $(T, V, \mu, \mathbf{v})$.

\subsubsection{Conjugate microcanonical ensemble $(\bar{L},V,N,\mathbf{v})$}
In the conjugate microcanonical ensemble $(\bar{L}, V, N, \mathbf{v})$, the thermodynamic potential $S$ and the quantities $T$, $p$, and $\mu$ are calculated by Eqs.~(\ref{ad40}) and (\ref{ad42}), respectively. Thermodynamic potential (\ref{ad40}) satisfies the relation $S = S_0$. By substituting Eq.~(\ref{44a}) and the relations $S = S_0$, $V = V_0/\gamma$, and $N = N_0$ into Eq.~(\ref{ad42}), and using Eq.~(\ref{ad12}), we obtain:
\begin{equation}\label{7ad}
  T=\frac{T_{0}}{\gamma}, \qquad p=p_{0}, \qquad \mu=\frac{\mu_{0}}{\gamma}.
\end{equation}
Thus, the Planck transformations for temperature, pressure, and chemical potential hold in the conjugate microcanonical ensemble $(\bar{L}, V, N, \mathbf{v})$.

\subsection{Velocity $\mathbf{v}$ as an independent state variable: Ott transformations}\label{sec5}
Consider the velocity $\mathbf{v}$ as an independent state variable. We derive the Ott transformations from the total energy $E(S, V, N, \mathbf{v})$ expressed in the state variables $(S, V, N, \mathbf{v})$, which is not a thermodynamic potential. According to thermodynamic principles, the appropriate thermodynamic potential for these variables is the function $\tilde{L}(S, V, N, \mathbf{v})$, obtained from the fundamental thermodynamic potential $E(S, V, N, \mathbf{P})$ through a Legendre transformation, or the conjugate fundamental potential $\bar{L}(S, V, N, \mathbf{v})$ (the negative of the Lagrangian function). In Hamiltonian mechanics, the total energy can not be a function of velocity as an independent state variable.

In the reference frame $K_{0}$, the fundamental thermodynamic potential is represented by the rest energy, as given by Eq.~(\ref{9a}). The Lorentz factor is determined by Eq.~(\ref{ad46}) and it is a function of the velocity $\mathbf{v}$. The thermodynamic state variables of this ensemble and the rest energy are transformed according to Eqs.~(\ref{10a}) and (\ref{ad43}). The rest volume $V_{0}$ at fixed values of $(V,\mathbf{v})$ is given by Eq.~(\ref{b1}). The total energy (\ref{6}) of the thermodynamic system $\mathcal{A}$ in the reference frame $K$ can be written as:
\begin{equation}\label{50a}
  E = \gamma(\mathbf{v}) E_{0}(S_{0},V_{0},N_{0}),
\end{equation}
where $S_{0}$, $V_{0}$, and $N_{0}$ are calculated by Eq.~(\ref{10a}). Differentiating $E$ with respect to the independent state variables $(S, V, N, \mathbf{v})$ of the thermodynamic system $\mathcal{A}$ in the reference frame $K$, and applying Eqs.~(\ref{6}), (\ref{9a}), (\ref{2}), (\ref{10a}), (\ref{ad46}), and (\ref{b1}), we obtain
\begin{align}\label{47z}
  dE &= \gamma T_{0}  dS - \gamma^{2} p_{0} dV + \gamma \mu_{0} dN \nonumber \\ &+ \gamma^{2} (1-\frac{p_{0}V_{0}}{E_{0}}) \ \mathbf{P} \cdot d\mathbf{v}.
\end{align}
The fundamental equation of thermodynamics in reference frame $K$ can be written as
\begin{equation}\label{47}
  dE = T_{*} dS - p_{*} dV + \mu_{*} dN + \mathbf{P}_{*} \cdot d\mathbf{v},
\end{equation}
where the thermodynamic quantities are defined as
\begin{align}\label{48}
 T_{*} &\equiv \left(\frac{\partial E}{\partial S}\right)_{VN\mathbf{v}}, \qquad p_{*} \equiv - \left(\frac{\partial E}{\partial V}\right)_{SN\mathbf{v}}, \nonumber \\
 \mu_{*} & \equiv \left(\frac{\partial E}{\partial N}\right)_{SV\mathbf{v}}, \qquad \mathbf{P}_{*} \equiv \left(\frac{\partial E}{\partial \mathbf{v}}\right)_{SVN}.
\end{align}
Here, $T_{*}$, $p_{*}$, $\mu_{*}$, and $\mathbf{P}_{*}$ represent the temperature, pressure, chemical potential, and generalized momentum, respectively, of the thermodynamic system $\mathcal{A}$ in the reference frame $K$.

By comparing Eqs.~(\ref{47}) and (\ref{47z}), we obtain the Ott transformations~\cite{Haar,Farias,Ott,Sutcliffe,Parvan2019} for the thermodynamic quantities defined in Eqs.~(\ref{48}) and (\ref{2}) (cf. Eq.~(\ref{39b})):
\begin{align}\label{39bb}
  T_{*}  &=  \gamma \ T_{0}, \qquad   p_{*} = \gamma^{2} \ p_{0}, \qquad
  \mu_{*} = \gamma \ \mu_{0}, \nonumber \\   \mathbf{P}_{*} &= \gamma^{2} (1-\frac{p_{0}V_{0}}{E_{0}})  \mathbf{P},
\end{align}
where the 3-momentum $\mathbf{P}$ is determined by Eq.~(\ref{44b}). See~\ref{App}.

The relation $p_{*} = \gamma^{2} p_{0}$ for pressure was first derived by Sutcliffe in~\cite{Sutcliffe}. The Ott transformations, as given by Eq.~(\ref{39bb}), are derived from the total energy $E(S, V, N, \mathbf{v})$, which is not a thermodynamic potential in the state variables $(S, V, N, \mathbf{v})$. Instead, the appropriate thermodynamic potentials for these variables are either $\tilde{L}(S, V, N, \mathbf{v})$ or $\bar{L}(S, V, N, \mathbf{v})$. In equilibrium thermodynamics, thermodynamic quantities are defined by the thermodynamic potential. Consequently, relativistic transformations of these quantities must also be derived from the corresponding thermodynamic potential. The transformation for heat transfer can be expressed as $\delta Q_{*}= \gamma \delta Q_{0}$~\cite{Haar,Farias,Nakamura}. Additionally, the ratio $p_{*} V/T_{*}=p_{0}V_{0}/T_{0}$ remains invariant~\cite{Nakamura,Sutcliffe}. By comparing Eq.~(\ref{39bb}) with Eq.~(\ref{39b}), we derive the relationship between the Ott and Planck thermodynamic quantities as (see~\ref{App})
\begin{equation}\label{s6}
  T_{*}=\gamma^{2} T, \qquad p_{*}=\gamma^{2} p, \qquad \mu_{*}=\gamma^{2} \mu.
\end{equation}
By comparing Eqs.~\eqref{41} and \eqref{47}, we derive the following relation for the state variables $(S, V, N, \mathbf{v})$:
\begin{equation}\label{s6d}
  dE=\gamma^{2} (d\bar{L}+2\mathbf{P} \cdot d\mathbf{v}).
\end{equation}
Consequently, the function $E(S, V, N, \mathbf{v})$ in the state variables $(S, V, N, \mathbf{v})$ does not qualify as a thermodynamic potential. It can be readily shown that, in the state variables $(S, V, N, \mathbf{v})$, the following equation holds:
\begin{equation}\label{n3}
  dE_{0} = \frac{1}{\gamma} (dE-\gamma^{2}\mathbf{P} \cdot d\mathbf{v}).
\end{equation}

\subsection{Summary}
Table~\ref{tab:1} presents the essential formulas for calculating thermodynamic quantities and the thermodynamic potentials of a moving relativistic thermodynamic system in the state variables $(S,V,N,\mathbf{P})$ and $(S,V,N,\mathbf{v})$.

The relativistic transformations for thermodynamic quantities are derived under the following conditions:
\begin{enumerate}
\item The thermodynamic system is at rest in a moving reference frame $K_0$, i.e., $\mathbf{P}_0 = 0$ or $\mathbf{P} = \mathbf{v}E$;
\item The reference frame $K_0$ is inertial, moving at a constant velocity $\mathbf{v}$ relative to the laboratory frame $K$;
\item The velocity $\mathbf{v}$ is an independent state variable;
\item The finite volume $V$ of the thermodynamic system;
\item The rest energy $E_0$ in $K_0$ is the internal energy of a thermodynamic system;
\item The internal energy $E_0$ of a thermodynamic system serves as the fundamental thermodynamical potential $E_{0}(S_{0},V_{0},N_{0})$ in the reference frame $K_0$, providing the sole connection between equilibrium thermodynamics and relativistic dynamics;
\item The total energy of the thermodynamic system in the laboratory reference frame $K$ is the fundamental thermodynamic potential $E(S, V, N, \mathbf{P})$ in the variables $(S,V,N,\mathbf{P})$ and coincides with the Hamiltonian;
\item The conjugate fundamental thermodynamic potential $\bar{L}(S, V, N, \mathbf{v})$ in the laboratory frame $K$, expressed in terms of the variables $(S, V, N, \mathbf{v})$, is defined by the negative Lagrangian function;
\item The total energy $E(S, V, N, \mathbf{v})$ of the thermodynamic system in the laboratory frame $K$, expressed in terms of $(S, V, N, \mathbf{v})$, is not a thermodynamic potential;
\item The entropy $S$ and particle number $N$ are invariant, while the volume $V$ undergoes Lorentz contraction.
\end{enumerate}

In this study, under these conditions, we obtained the following results:
\begin{enumerate}
  \item The Planck transformations are derived from the conjugate fundamental potential $\bar{L}(S, V, N, \mathbf{v})$ in the state variables
  $(S, V, N,\mathbf{v})$ in the laboratory reference frame $K$;
  \item The Ott transformations are derived from the total energy $E(S, V, N, \mathbf{v})$ expressed in terms of the state variables $(S, V, N, \mathbf{v})$ in the laboratory reference frame $K$, which is not a thermodynamic potential;
  \item The fundamental potential $E(S, V, N, \mathbf{P})$, expressed in terms of the state variables $(S, V, N, \mathbf{P})$ in the laboratory reference frame $K$, yields the Non-Planck transformations, which differ from the Planck transformations by a factor $\alpha$. In this framework, the Lorentz factor $\gamma$ depends on the thermodynamic variables, a relationship that lacks experimental validation;
  \item The Planck transformations remain valid in other ensembles corresponding to conjugate potentials derived from the conjugate fundamental potential $\bar{L}$ through Legendre transformations or by substituting variable $S$ with $\bar{L}$: $\bar{F}(T, V, N, \mathbf{v})$, $\bar{G}(T, p, N, \mathbf{v})$, $\bar{\Omega}(T, V, \mu, \mathbf{v})$ and $S(\bar{L}, V, N, \mathbf{v})$;
  \item The thermodynamic potential $\tilde{L}(S, V, N, \mathbf{v})$, defined for the state variables $(S, V, N, \mathbf{v})$ and derived from the fundamental potential $E(S, V, N, \mathbf{P})$ via Legendre transformation, differs from the conjugate fundamental potential $\bar{L}(S, V, N, \mathbf{v})$ by a factor $\alpha$;
  \item The thermodynamic potential $\bar{E}(S, V, N, \mathbf{P})$, defined for the state variables $(S, V, N, \mathbf{P})$ and derived from the conjugate fundamental thermodynamic potential $\bar{L}(S, V, N, \mathbf{v})$ through a Legendre transformation, is distinct from the fundamental thermodynamic potential $E(S, V, N, \mathbf{P})$ and precisely aligns with the formula from Tolman~\cite{Tolman}.
\end{enumerate}

The primary distinction between Ott and Planck transformations lies in the choice of thermodynamic potential for the state variables $(S, V, N, \mathbf{v})$. Both transformations utilize the same state variables $(S, V, N, \mathbf{v})$ and Lorentz factor $\gamma$, but they differ in their thermodynamic potentials (see Table~\ref{tab:1}, columns 2 and 3). Planck transformations are derived from the conjugate fundamental thermodynamic potential, whereas Ott transformations are based on the function $E(S, V, N, \mathbf{v})$, which is not a thermodynamic potential for these state variables.

When momentum $\mathbf{P}$ is treated as a state variable, thermodynamic principles require that the Lorentz factor $\gamma$ and velocity $\mathbf{v}$ depend on the thermodynamic state variables, as derived from the self-consistency equation (\ref{ad44}). In contrast, when velocity $\mathbf{v}$ is an independent state variable, the Lorentz factor $\gamma$ depends solely on velocity $\mathbf{v}$ and is independent of thermodynamic variables.

The rest energy $ E_0 $ given by Eq.~(\ref{9a}), representing the internal energy of a thermodynamic system, is the key quantity linking equilibrium thermodynamics to relativistic mechanics. In the canonical ensemble, $ E_0 $ is a function of ($ T_0 $, $ V_0 $, $ N_0 $); in the grand canonical ensemble, it depends on ($ T_0 $, $ V_0 $, $ \mu_0 $); and in the fundamental ensemble, it is a function of ($ S_0 $, $ V_0 $, $ N_0 $). Equilibrium thermodynamics mandates the dependence described by equation (8) in the fundamental ensemble, which is not an assumption but a requirement for consistency with relativistic mechanics. In equilibrium thermodynamics, the internal energy $E_0$ is the unique quantity linking thermodynamics with relativistic mechanics, serving as the essential bridge between these fundamentally distinct theories.

The present formalism begins with the 4-momenta, $ P^\mu $ and $ P^{\mu}_0 $, of a thermodynamic system in reference frames $K$ and $K_0$, respectively. The components of these 4-momenta are related by Lorentz transformations. We require the thermodynamic system to be at rest in reference frame $K_0$, such that its 3-momentum, $\mathbf{P}_0 = 0$. This condition simplifies the Lorentz transformations for the 4-momentum components to the form given in Eq.~(\ref{6}), and the system's 3-velocity equals the velocity of $K_0$ relative to $K$. In this case, the zeroth component of the 4-momentum in the $K_0$ reference frame, $ P^0_0 $, represents the rest energy, which corresponds to the internal energy of the thermodynamic system. This equivalence enables the integration of thermodynamics into the relativistic formalism by defining $ P^0_0 $ as a function of the thermodynamic state variables, adopting it as the fundamental thermodynamic potential in $K_0$. We next consider two possibilities for the independent state variable of the thermodynamic system: its 3-momentum or its 3-velocity. If we choose the 3-momentum, denoted as $ \mathbf{P} $, in the reference frame $ K $ as the independent state variable, the zeroth component of the 4-momentum, $ P^0 $, emerges as a fundamental thermodynamic potential. This potential depends on $ \mathbf{P} $ and the thermodynamic state variables. In this scenario, the zeroth component $ P^0 $ of the thermodynamic system's 4-momentum corresponds to the Hamiltonian of a freely moving particle without interactions (see Eq. (9.7) in Ref.~\cite{Landau}). This Hamiltonian is determined using the Lorentz transformation or, alternatively, derived from the invariance of the four-momentum squared, $P_\mu P^\mu = E_0^2$. The energy expressed as a function of momentum is called the Hamiltonian function~\cite{Landau}. Alternatively, if we select the 3-velocity $ \mathbf{v} $ as the independent state variable, the principles of mechanics necessitate replacing the zeroth component $ P^0 $ of the 4-momentum with a function obtained via the Legendre transformation, known as the Lagrangian. This Lagrangian precisely corresponds to that of a freely moving particle without interactions, which can be derived from the action for a single particle with three-velocity $\mathbf{v}$ and rest energy $E_0$ as given in Eq.~(8.2) of Ref.~\cite{Landau}. We define the negative of this Lagrangian, denoted as $ \bar{L} $, as the conjugate fundamental thermodynamic potential, which depends on $ \mathbf{v} $ and the thermodynamic state variables. The thermodynamic state variables in the zeroth component $ P^0 $ of the 4-momentum transform according to equation~(\ref{10a}). When the 3-momentum $ \mathbf{P} $ is chosen as the independent variable, the Lorentz factor depends on the thermodynamic state variables. In contrast, when the 3-velocity $ \mathbf{v} $ is chosen as the independent variable, the Lorentz factor remains independent of these variables. We next define the thermodynamic quantities using the standard approach of equilibrium thermodynamics. This involves differentiating the fundamental thermodynamic potential and its conjugate with respect to the independent state variables in the reference frame $ K $. From the fundamental thermodynamic potential $ P^0 $, with the 3-momentum $ \mathbf{P} $ as the independent state variable, we derive the Non-Planck transformations for the thermodynamic quantities. Conversely, from the conjugate fundamental thermodynamic potential $ \bar{L} $, with the 3-velocity $ \mathbf{v} $ as the independent state variable, we obtain the Planck transformations. However, if we define the thermodynamic quantities from the Hamiltonian $ P^0 $ using $ \mathbf{v} $ as the independent state variable, in violation of the principles of mechanics, we derive the Ott transformations for the thermodynamic quantities.

In our formalism, we assign a four-momentum to a thermodynamic system in a manner analogous to that employed in relativistic nuclear physics (see, e.g., Ref.~\cite{Baldin}). This approach remains consistent, as the thermodynamic systems considered have infinitesimal dimensions on the order of $\sim 1$ fm. In this formalism, a moving thermodynamic system is defined to be in thermodynamic equilibrium by construction. In relativistic thermodynamics, attempts to derive the zeroth law of thermodynamics from first principles have indeed been explored in the work of Gavassino~\cite{Gavassino}. According to Einstein's principle of relativity, the fact that a system is moving at a constant velocity has no effect on its thermodynamic equilibrium. Whether the system is in motion or at rest, its thermodynamic equilibrium remains unchanged. This principle is fundamental to special relativity. Note that the paper~\cite{Gavassino} overlooks the fact that relativistic nuclear systems can interact with each other via gamma photons, resulting in recoil and alterations in the momentum and velocity of the nuclei (thermodynamic systems). Consequently, this proof method is incompatible with the principles of special relativity.

\section{Thermodynamics of a moving quark-gluon plasma system}\label{sec6}
In parton physics, the proton can be viewed as a composite system of valence quarks, gluons, and a sea of quark-antiquark pairs~\cite{Halzen}. Consequently, it is often modeled as a thermodynamic system defined by thermodynamic quantities such as temperature and pressure~\cite{Bourrely2002,Soffer2019,Burkert2018,Mac1989}. However, treating a relativistically moving proton as a thermodynamic system poses the challenge of transforming its thermodynamic quantities across different reference frames.

Consider an ultrarelativistic ideal gas of quarks and gluons with $N_f = 2+1$ flavors, described within the grand canonical ensemble at a chemical potential of $\mu_0 = 0$ (the Stefan–Boltzmann limit). In this case, the thermodynamic potential, energy, entropy, and pressure of the thermodynamic system $\mathcal{A}$, at rest in the inertial reference frame $K_0$, can be expressed as detailed in~\cite{Yagi2008}
\begin{align}\label{53}
  \Omega_{0} &= -\frac{\pi^{2}gV_{0}}{90}  T_{0}^{4}, \qquad    E_{0} = \frac{\pi^{2}gV_{0}}{30}  T_{0}^{4}, \\ \label{54}
   S_{0} &= \frac{2\pi^{2}gV_{0}}{45}  T_{0}^{3}, \qquad    p_{0} = \frac{\pi^{2}g}{90}  T_{0}^{4},
\end{align}
where $g$ represents the effective degeneracy factor, which is defined as~\cite{Yagi2008,Brown,Parvan2020}
\begin{align}\label{55}
  g &= g_{g}+\frac{7}{8}  g_{q}, \quad  g_{g} = 2_{spin} \times (N_{c}^{2}-1), \nonumber \\
  g_{q} &= 2_{spin} \times 2_{q\bar{q}} \times N_{c} \times N_{f}.
\end{align}

Now, let us examine the fundamental thermodynamic ensemble in the reference frame $K_{0}$. The fundamental thermodynamic potential $E_{0}$ is derived from the grand thermodynamic potential (see Eq.~(\ref{53})) through an inverse Legendre transformation:
\begin{equation}\label{55a}
  E_{0} = \Omega_{0} +T_{0}S_{0}+\mu_{0}N_{0}.
\end{equation}
Given that $\mu_0 = 0$, the fundamental thermodynamic potential (Eq.~(\ref{55a})) and the thermodynamic quantities (Eq.~(\ref{2})) of the thermodynamic system $\mathcal{A}$, at rest in the reference frame $K_0$, can be expressed as follows:
\begin{align}\label{56}
  E_{0} &=  \frac{3}{4} a \ \frac{S_{0}^{4/3}}{V_{0}^{1/3}}, \qquad a = \left(\frac{45}{2\pi^{2}g} \right)^{1/3}, \\   \label{57}
  T_{0} &= a \ \frac{S_{0}^{1/3}}{V_{0}^{1/3}}, \\ \label{58}
  p_{0} &= \frac{a}{4} \ \frac{S_{0}^{4/3}}{V_{0}^{4/3}}.
\end{align}

Let us examine a thermodynamic system of quarks and gluons that is in motion relative to the laboratory reference frame $K$, while staying stationary within the moving reference frame $K_{0}$.

\subsection{Fundamental ensemble $(S,V,\mathbf{P})$}
The independent state variables in the fundamental ensemble are $(S,V,\mathbf{P})$. Substituting Eq.~(\ref{10a}) into (\ref{56})--(\ref{58}), we obtain
\begin{align}\label{x10}
  E_{0} &= \frac{3}{4} \frac{a}{\gamma^{1/3}}  \frac{S^{4/3}}{V^{1/3}}, \\ \label{x11}
  T_{0} &= \frac{a}{\gamma^{1/3}} \frac{S^{1/3}}{V^{1/3}}, \\ \label{x12}
  p_{0} &= \frac{1}{4}\frac{a}{\gamma^{4/3}} \ \frac{S^{4/3}}{V^{4/3}}.
\end{align}
Substituting Eqs.~(\ref{x10})--(\ref{x12}) into Eqs.~(\ref{23a}) and (\ref{a7}), we have
\begin{align} \label{x13}
  E &= \gamma E_{0} = \frac{3}{4} a \gamma^{2/3}  \frac{S^{4/3}}{V^{1/3}}, \\ \label{x20}
  T &=  \frac{\alpha T_{0}}{\gamma} = \frac{\alpha a}{\gamma^{4/3}} \frac{S^{1/3}}{V^{1/3}}, \\ \label{x21}
  p &=  \alpha p_{0} = \frac{1}{4}\frac{\alpha a}{\gamma^{4/3}} \ \frac{S^{4/3}}{V^{4/3}}.
\end{align}
Substituting Eq.~(\ref{x10}) into Eq.~(\ref{11}) or Eq.~(\ref{ad44}), we obtain the equation for the unknown $\gamma$ as
\begin{equation}\label{x6}
  \gamma^{2} - B V^{2/3} \gamma^{2/3} - 1 = 0,
\end{equation}
where
\begin{equation}\label{x6a}
  B \equiv \frac{16}{9} \frac{\mathbf{P}^{2}}{a^{2} S^{8/3}}.
\end{equation}
This equation can be simplified into a cubic equation through the substitution $y=\gamma^{2/3}$. The solution to this cubic equation is then given by Cardano's formula:
\begin{align}\label{x7}
  \gamma &=   \left(F_{+}^{1/3} + F_{-}^{1/3} \right)^{3/2}, \\ \label{x8}
  F_{\pm} &= \frac{1}{2} \pm  \sqrt{\frac{1}{4}-\frac{1}{27}B^{3}V^{2}}.
\end{align}
The functions $F_{\pm}$ satisfy the relations:
\begin{align}\label{x7a}
  F_{+} F_{-} &= \frac{1}{27}B^{3}V^{2}, \\ \label{x7b}
  F_{+} + F_{-} &= 1,  \\ \label{x7c}
  F_{+} - F_{-} &= (F_{+}^{1/3} - F_{-}^{1/3}) [\gamma^{4/3}-(F_{+} F_{-})^{1/3}]  \nonumber \\
  &= 2 \sqrt{\frac{1}{4}-\frac{1}{27}B^{3}V^{2}}.
\end{align}
The rest volume $V_{0}$ at fixed values of $(S,V,\mathbf{P})$ is given by
\begin{equation}\label{x15}
  V_{0} = \left(F_{+}^{1/3} + F_{-}^{1/3} \right)^{3/2} V.
\end{equation}
The function $\alpha$, which is defined in Eq.~(\ref{23}), can be expressed as
\begin{equation} \label{x18}
  \alpha = \gamma^{4/3} \ \frac{F_{+}^{1/3} - F_{-}^{1/3}}{F_{+} - F_{-}},
\end{equation}
where $V_{0}p_{0}/E_{0}=1/3$. Taking the partial derivatives of the fundamental thermodynamic potential $E$, given by Eq.~(\ref{x13}), with respect to the state variables $(S,V)$ and using Eqs.~(\ref{x18}) and (\ref{a7}), we obtain
\begin{align}\label{x19a}
  T &=\frac{\partial E}{\partial S} =   \frac{a S^{1/3}}{V^{1/3}} \ \frac{F_{+}^{1/3} - F_{-}^{1/3}}{F_{+} - F_{-}} = \frac{\alpha a}{\gamma^{4/3}} \frac{S^{1/3}}{V^{1/3}}, \\ \label{x20a}
 p &=  -\frac{\partial E}{\partial V} =  \frac{a S^{4/3}}{4 V^{4/3}} \ \frac{F_{+}^{1/3} - F_{-}^{1/3}}{F_{+} - F_{-}} = \frac{\alpha a}{4\gamma^{4/3}}  \frac{S^{4/3}}{V^{4/3}}.
\end{align}
Thus, the temperature (\ref{x20}) and pressure (\ref{x21}) satisfy the Non-Planck transformations (Eq.~(\ref{a7})).

\subsection{Planck transformations in conjugate fundamental ensemble $(S,V,\mathbf{v})$}
The independent state variables in the conjugate fundamental ensemble are $(S,V,\mathbf{v})$. Substituting Eq.~(\ref{10a}) into (\ref{56})--(\ref{58}), we obtain Eqs.~(\ref{x10})--(\ref{x12}). Substituting Eq.~(\ref{x10}) into Eq.~(\ref{44a}), we get the conjugate fundamental thermodynamic potential for an ideal gas in the reference frame $K$:
\begin{equation} \label{p7}
  \bar{L} =\frac{3}{4} \frac{a}{\gamma^{4/3}}  \frac{S^{4/3}}{V^{1/3}},
\end{equation}
where the Lorentz factor $\gamma$ is calculated by Eq.~(\ref{ad46}). Substituting Eq.~(\ref{p7}) into Eq.~(\ref{42}) and using Eqs.~(\ref{ad46}), (\ref{x11}), and (\ref{x12}), we obtain the temperature and pressure in the conjugate fundamental ensemble:
\begin{align}\label{p8}
  T &=  \frac{a}{\gamma^{4/3}} \ \frac{S^{1/3}}{V^{1/3}} = \frac{T_{0}}{\gamma}, \\ \label{p9}
  p &=  \frac{1}{4} \ \frac{a}{\gamma^{4/3}} \ \frac{S^{4/3}}{V^{4/3}} = p_{0}.
\end{align}
The temperature (\ref{p8}) and pressure (\ref{p9}) conform to the Planck transformations given in Eq.~(\ref{39b}).

\subsection{Ott transformations in state variables $(S,V,\mathbf{v})$}
The independent state variables are $(S,V,\mathbf{v})$. Substituting Eq.~(\ref{10a}) into (\ref{56})--(\ref{58}), we obtain Eqs.~(\ref{x10})--(\ref{x12}). Substituting Eq.~(\ref{x10}) into Eq.~(\ref{50a}), we obtain the thermodynamic potential as function of velocity $\mathbf{v}$ (cf.~Eq.~(\ref{x13})):
\begin{equation} \label{x13a}
  E =  \frac{3}{4} a \gamma^{2/3}  \frac{S^{4/3}}{V^{1/3}}.
\end{equation}
Substituting Eq.~(\ref{x13a}) into Eq.~(\ref{48}) and using Eqs.~(\ref{x11}) and (\ref{x12}), we derive
\begin{align}\label{p14}
  T_{*} &=  a \gamma^{2/3} \ \frac{S^{1/3}}{V^{1/3}} = \gamma T_{0}, \\ \label{p15}
  p_{*} &=  \frac{1}{4} a \gamma^{2/3} \ \frac{S^{4/3}}{V^{4/3}} = \gamma^{2} p_{0},
\end{align}
where the Lorentz factor $\gamma$ is calculated by Eq.~(\ref{ad46}). The temperature (\ref{p14}) and pressure (\ref{p15}) conform to the Ott transformations~(\ref{39bb}).

\begin{figure}[!htb]
%\vspace*{0.0cm}
\minipage{0.49\textwidth}
\includegraphics[width=\linewidth]{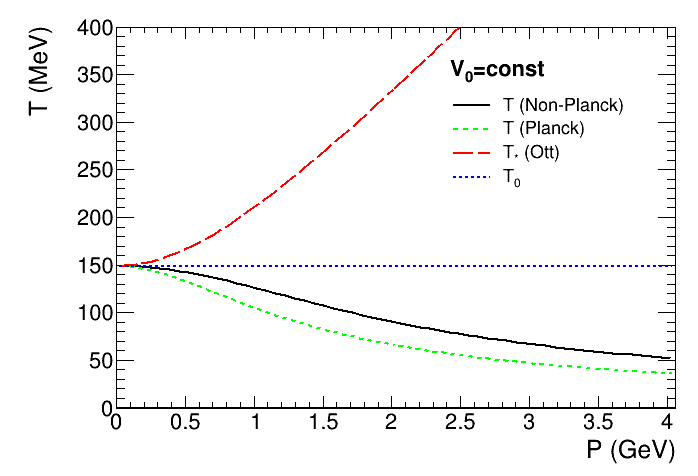}
\endminipage\hfill
\minipage{0.49\textwidth}
\vspace*{-0.2cm}
%\hspace*{0cm}
\includegraphics[width=\linewidth]{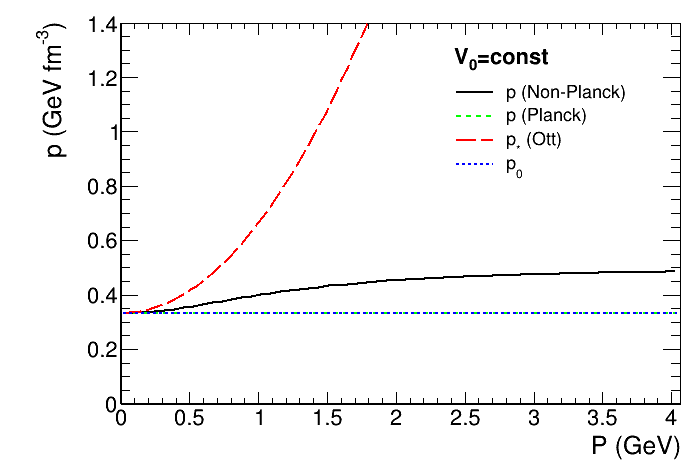}
\endminipage\hfill
\caption{(Color online) The temperature $T$ and pressure $p$ in the laboratory reference frame $K$ are expressed as functions of the system's momentum $\mathbf{P}$, where the system is at rest in the moving reference frame $K_{0}$. This pertains to an ultrarelativistic ideal gas of quarks and gluons with $N_{f} = 2+1$ flavors in a system of mass $E_{0}=1$ GeV in a volume $V_{0} = 1$ fm$^{3}$ at $\mu_{0} = 0$. The temperatures $T$ (Non-Planck), $T$ (Planck), $T_{*}$, and $T_{0}$ were computed using Eqs.~(\ref{x20}), (\ref{p8}), (\ref{p14}), and (\ref{57}), respectively. Similarly, the pressures $p$ (Non-Planck), $p$ (Planck), $p_{*}$, and $p_{0}$ were determined using Eqs.~(\ref{x21}), (\ref{p9}), (\ref{p15}), and (\ref{58}), respectively.}
\label{fig1}
\end{figure}

\subsection{Numerical results}
Figure~\ref{fig1} illustrates the behavior of temperature and pressure in the laboratory reference frame $K$ as functions of the system's momentum $\mathbf{P}$ for an ultrarelativistic ideal gas of quarks and gluons with $N_{f} = 2+1$ flavors in the system of mass $E_{0}=1$ GeV in a volume $V_{0} = 1$ fm$^{3}$ at $\mu_{0} = 0$. In the reference frame $K_{0}$, the system's entropy is $S_{0} = 8.95$, with a temperature of $T_{0} = 148.9$ MeV and a pressure of $p_{0} = 0.333$ GeV/fm$^{3}$. The curves $T$ (Non-Planck), $T$ (Planck), $T_{*}$ (Ott), and $T_{0}$ represent temperature calculations for the ideal gas based on Eqs.~(\ref{x20}), (\ref{p8}), (\ref{p14}), and (\ref{57}), respectively. Similarly, the curves $p$ (Non-Planck), $p$ (Planck), $p_{*}$ (Ott), and $p_{0}$ depict pressure calculations derived from Eqs.~(\ref{x21}), (\ref{p9}), (\ref{p15}), and (\ref{58}), respectively.

The temperatures $T$ (Non-Planck) and $T$ (Planck) exhibit similar behavior, decreasing as the momentum $\mathbf{P}$ increases and approaching zero as $\mathbf{P} \to \infty$. In contrast, the Ott temperature $T_{*}$ increases with $\mathbf{P}$ and, unlike the Planck temperature, diverges to infinity as $\mathbf{P} \to \infty$. The Planck pressure $p$ (Planck) and the rest pressure $p_0$ remain constant, independent of the system's momentum $\mathbf{P}$. However, the Ott pressure $p_{*}$ increases with $\mathbf{P}$ and, in contrast to the Planck pressure, diverges to infinity as $\mathbf{P} \to \infty$. The Non-Planck pressure $p$ (Non-Planck) also increases with $\mathbf{P}$, but approaches a constant value, $p = \frac{3}{2} p_0$, as $\mathbf{P} \to \infty$.

\begin{figure}[!htb]
\includegraphics[width=0.49\textwidth]{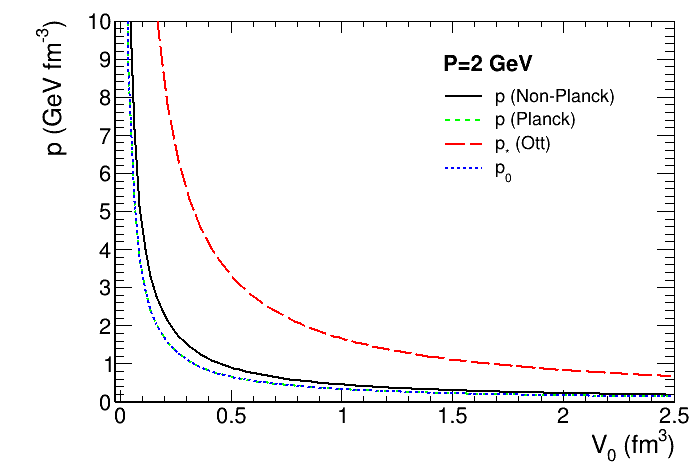}
\caption{(Color online) The equation of state for an ultrarelativistic ideal gas of quarks and gluons with $N_{f} = 2+1$ flavors in a system of mass $E_{0}=1$ GeV with $P = 2$ GeV at $\mu_{0} = 0$. For additional details, refer to the caption of Figure~\ref{fig1}.}
\label{fig2}
\end{figure}

Figure~\ref{fig2} depicts the behavior of pressure in the laboratory reference frame $K$ as a function of the rest volume $V_{0}$ for an ultrarelativistic ideal gas of quarks and gluons with $N_{f} = 2+1$ flavors in the system of mass $E_{0}=1$ GeV for $P = 2$ GeV at $\mu_{0} = 0$. The curves $p$ (Non-Planck), $p$ (Planck), $p_{*}$ (Ott), and $p_{0}$ represent pressure calculations derived from Eqs.~(\ref{x21}), (\ref{p9}), (\ref{p15}), and (\ref{58}), respectively. The Planck pressure $p$ (Planck), along with the rest pressure $p_{0}$, are identical. All pressures -- $p$ (Non-Planck), $p$ (Planck), $p_{*}$, and $p_{0}$ -- exhibit a rapid decrease as $V_{0}$ increases. Notably, the Ott pressure $p_{*}$ exceeds the Non-Planck pressure $p$ (Non-Planck) and Planck pressure $p$ (Planck) as well as the rest pressure $p_{0}$.

In the reference frame $K$, the energy density is given by $\varepsilon = \gamma^2 \varepsilon_0$, where $\varepsilon_0 = E_0 / V_0$ represents the energy density in the reference frame $K_0$. The Planck pressure in reference frame $K$ is expressed as $p = \gamma^{-2} \varepsilon / 3$. In contrast, the Ott-Sutcliffe pressure in reference frame $K$ is $p_* = \varepsilon / 3$, with its trace anomaly $(\varepsilon - 3 p_*) / T_{*}^{4} = 0$, consistent with the Stefan–Boltzmann equation. However, the Planck pressure $p$ is inconsistent with the Stefan–Boltzmann equation, as its trace anomaly is non-zero.

\section{Boltzmann-Gibbs and Tsallis blast-wave models for finite-volume fireballs}
A comparison of hadron transverse momentum distributions, as predicted by the ordinary statistical model and the blast-wave model, can elucidate which of the Planck or Ott transformations is more accurate.

In the conventional Boltzmann-Gibbs or Tsallis statistical framework, the transverse momentum distribution reflects a globally equilibrated fireball -- a thermodynamic system with volume $V$ and temperature $T$ -- stationary in the laboratory frame $K$. In the grand canonical ensemble, the Maxwell-Boltzmann global equilibrium transverse momentum distribution of hadrons, based on Boltzmann-Gibbs statistics, is expressed as~\cite{Parvan2017}:
\begin{equation}\label{t1}
\frac{d^{2}N}{dp_{T}dy} = \frac{gV}{(2\pi)^{2}} p_{T}  m_{T} \cosh y  \ e^{-\frac{m_{T} \cosh y-\mu}{T}},
\end{equation}
where $m_{T}=\sqrt{p_{T}^{2}+m^{2}}$ is the transverse mass, $p_{T}$ and $y$ are the transverse momentum and rapidity, respectively, $V$ is the volume, $T$ and $\mu$ are the temperature and chemical potential in the laboratory reference frame $K$, and $g$ is the spin degeneracy factor.

The Maxwell-Boltzmann global equilibrium transverse momentum distribution of hadrons, derived from Tsallis-3 statistics in the zeroth-order approximation, can be written as~\cite{Cleymans12a,Cleymans2012,Parvan2023}
\begin{align}\label{t2}
\frac{d^{2}N}{dp_{T}dy} &= \frac{gV}{(2\pi)^{2}} p_{T}  m_{T} \cosh y  \nonumber \\
                        & \times \left[1-(1-q) \frac{m_{T} \cosh y-\mu}{T} \right]^{\frac{q}{1-q}},
\end{align}
where $q\in\mathbb{R}$ is the non-extensivity parameter, taking values in the range $0<q<\infty$. The Tsallis global equilibrium transverse momentum distribution (Eq.~(\ref{t2})) is an equilibrium distribution, as the Tsallis statistics~\cite{Tsal88,Tsal98} is derived from the maximum entropy principle, which defines the system's equilibrium state.

\subsection{Boltzmann-Gibbs blast-wave model for finite-volume fireballs}
The blast-wave model is established based on the principle of local equilibrium. The Maxwell-Boltzmann local equilibrium transverse momentum distribution of hadrons in the Boltzmann-Gibbs blast-wave model can be written as~\cite{Florkowski}
\begin{equation}\label{t3}
\frac{d^{2}N}{dp_{T}dy} = \frac{g}{(2\pi)^{3}} p_{T} \int d\Sigma_{\mu} p^{\mu}   \int\limits_{0}^{2\pi} d\varphi \ e^{-\frac{p_{\mu}u^{\mu}-\mu_{0}}{T_{0}}}.
\end{equation}
Here, $d\Sigma_{\mu}$ represents the volume element of the freeze-out hypersurface, $p^{\mu} = (E_{\mathbf{p}}, \mathbf{p})$ denotes the particle's four-momentum, and $u^{\mu} = \gamma (1, \mathbf{v})$ is the four-velocity of the fluid element (thermodynamic system). The variables $p_{T}$, $y$, and $\varphi$ represent the transverse momentum, rapidity, and azimuthal angle of a particle, respectively. Additionally, $T_{0}$ and $\mu_{0}$ are the temperature and chemical potential of the fluid element at rest, measured in the reference frame $K_{0}$, which moves at a constant velocity $\mathbf{v}$ relative to the laboratory frame $K$.

In this paper, we focus exclusively on the cylindrically symmetric hypersurface, which, within the framework of Boltzmann-Gibbs statistics, aligns with the blast - wave model proposed by Schnedermann, Sollfrank, and Heinz~\cite{Schnedermann}. By adopting the computational approach outlined in Florkowski~\cite{Florkowski}, we derive the Maxwell - Boltzmann local equilibrium transverse momentum distribution as follows:
\begin{align}\label{t3a}
& \frac{d^{2}N}{dp_{T}dy} = \frac{g e^{\frac{\mu_{0}}{T_{0}}}}{(2\pi)^{3}} \ p_{T}  m_{T}   \nonumber \\
& \;\;  \tau_{\mathrm{f}} \int\limits_{0}^{R} r dr \int\limits_{-\eta_{max}}^{\eta_{max}} d\eta_{\|} \cosh(\eta_{\|}-y) \int\limits_{0}^{2\pi} d\phi \int\limits_{0}^{2\pi} d\varphi  \nonumber \\
& \;\;  e^{- \frac{1}{T_{0}}\left\{m_{T}\cosh\rho(r)\cosh(\eta_{\|}-y) - p_{T} \sinh\rho(r)\cos(\phi-\varphi)\right\}},
\end{align}
where
\begin{align}\label{t5}
  \cosh\rho(r) &=\frac{1}{\sqrt{1-\bar{v}_T^{2}(r)}}, \nonumber  \\
  \sinh\rho(r) &=\frac{\bar{v}_T(r)}{\sqrt{1-\bar{v}_T^{2}(r)}}.
\end{align}
In this context, $\bar{v}_T(r)$ denotes the transverse velocity of a fluid element in the plane $z=0$, with $\bar{v}_{T\mathrm{f}}$ representing its maximum value at $r=R$:
\begin{equation}\label{t6}
  \bar{v}_T(r) = \bar{v}_{T\mathrm{f}} \ \frac{r}{R}, \qquad \bar{v}_{T\mathrm{f}} \equiv \frac{R}{\tau_{\mathrm{f}}},
\end{equation}
where $R$ is the radius of the freeze-out firecylinder, and $\tau_{\mathrm{f}}$ is the fixed proper time at freeze-out. Here, $r$, $\eta_{\|}$ and $\phi$ represent the radial distance from the $z$-axis, the spacetime rapidity, and the azimuthal angle of the fluid element, respectively. Integrating Eq.~(\ref{t3a}) over the azimuthal angles $\phi$ and $\varphi$ and applying the second relation from Eq.~(\ref{t6}), we obtain
\begin{align}\label{t4}
\frac{d^{2}N}{dp_{T}dy} &= \frac{g e^{\frac{\mu_{0}}{T_{0}}}}{2\pi} \ p_{T}  m_{T} \frac{R}{\bar{v}_{T\mathrm{f}}} \int\limits_{0}^{R} r dr \int\limits_{-\eta_{max}}^{\eta_{max}} d\eta_{\|} \nonumber \\
                        & \times \cosh(\eta_{\|}-y) \ e^{-\frac{m_{T}}{T_{0}} \cosh\rho(r)\cosh(\eta_{\|}-y)} \nonumber \\
                        & \times I_{0}\left(\frac{p_{T}}{T_{0}} \sinh\rho(r) \right),
\end{align}
where $I_{0}(x)$ is the modified Bessel function of the first kind.

The volume of the firecylinder at the freeze-out stage can be expressed as
\begin{align}\label{t7}
  V &\equiv \int d\Sigma^{0} \nonumber \\
                            & =  \int\limits_{0}^{R} r dr  \int\limits_{-\eta_{max}}^{\eta_{max}} d\eta_{\|} \cosh\eta_{\|} \int\limits_{0}^{2\pi} d\phi \ \tau_{\mathrm{f}} \nonumber \\
                            &  = 2 z_{max} \pi R^{2}, \\  \label{t8}
                  z_{max} &= \tau_{\mathrm{f}} \sinh\eta_{max} = \frac{R}{\bar{v}_{T\mathrm{f}}} \sinh\eta_{max}.
\end{align}
Here, the definition of $\Sigma^{0}$ for a cylindrically symmetric hypersurface is detailed in Ref.~\cite{Florkowski}. Thus, Eq.~(\ref{t4}) describes the Maxwell-Boltzmann local equilibrium transverse momentum distribution for the Boltzmann-Gibbs blast-wave model within the finite volume of the firecylinder at the freeze-out stage.

Let us consider the maximal spacetime rapidity of the firecylinder $\eta_{max} = \infty$. Under this condition, the length of the firecylinder at freeze-out becomes infinite, with $z_{max} = \infty$, resulting in an infinite volume as well. By substituting $\eta_{max} = \infty$ into Eq.~(\ref{t4}), we derive the well-known formula for the transverse momentum distribution of the blast-wave model as~\cite{Schnedermann}
\begin{align}\label{t9}
\frac{d^{2}N}{dp_{T}dy} &= \frac{g e^{\frac{\mu_{0}}{T_{0}}}}{\pi} \ p_{T}  m_{T} \tau_{\mathrm{f}} \int\limits_{0}^{R} r dr K_{1}\left(\frac{m_{T}}{T_{0}} \cosh\rho(r)\right)  \nonumber \\
                        & \times  I_{0}\left(\frac{p_{T}}{T_{0}} \sinh\rho(r) \right).
\end{align}
Furthermore, by implementing additional simplifications and assuming $\bar{v}_T(r) = \text{const}$, we derive
\begin{align}\label{t9a}
\frac{d^{2}N}{dp_{T}dy} &= \frac{g e^{\frac{\mu_{0}}{T_{0}}}}{2\pi} \ p_{T}  m_{T} \tau_{\mathrm{f}} R^{2} \ K_{1}\left(\frac{m_{T}}{T_{0}\sqrt{1-\bar{v}_T^{2}}}\right)  \nonumber \\
                        & \times  I_{0}\left(\frac{p_{T} \bar{v}_T}{T_{0}\sqrt{1-\bar{v}_T^{2}}} \right).
\end{align}
Equation~(\ref{t9a}) represents one of the most widely utilized distributions in relativistic heavy-ion research (see Ref.~\cite{Florkowski} and the references cited therein).
However, the distributions (\ref{t9}) and (\ref{t9a}) pertain to the exceptional case of an infinite volume $V = \infty$ for the firecylinder at freeze-out, which is quite unusual in heavy ion collisions.

In the blast-wave model, the fluid element (thermodynamic system) moves with velocity $\mathbf{v}$ in the laboratory reference frame $K$, while remaining at rest in the moving reference frame $K_{0}$. The transverse momentum distributions (\ref{t3}), (\ref{t4}), (\ref{t9}), and (\ref{t9a}) are provided by the rest temperature $T_{0}$ and rest chemical potential $\mu_{0}$ of the fluid element in the reference frame $K_{0}$. Consequently, the temperature $T$ and chemical potential $\mu$ of the fluid element in the laboratory reference frame $K$ can be determined using the Planck or Ott transformations. By applying the Planck transformations (\ref{39b}) and the Ott transformations (\ref{39bb}), we can rewrite the distribution (\ref{t3}) as follows:
\begin{equation}\label{t10}
\frac{d^{2}N}{dp_{T}dy} = \frac{g}{(2\pi)^{3}} p_{T} \int d\Sigma_{\mu} p^{\mu}   \int\limits_{0}^{2\pi} d\varphi \ e^{-\frac{\gamma^{l} p_{\mu}u^{\mu}-\mu}{T}},
\end{equation}
where $l = -1$ corresponds to the Planck transformations, while $l = 1$ applies to the Ott transformations. The Lorentz factor, expressed in cylindrical parametrization, is given by~\cite{Florkowski}
\begin{equation}\label{t11}
  \gamma(r,\eta_{\|}) = \cosh\rho(r) \cosh\eta_{\|}.
\end{equation}
Substituting Eq.~(\ref{t11}) into Eq.~(\ref{t10}) and applying the same derivation method used for Eq.~(\ref{t4}), we derive the Maxwell-Boltzmann local equilibrium transverse momentum distribution for the Boltzmann-Gibbs blast-wave model using the Planck transformation ($l = -1$), expressed as
\begin{align}\label{t12}
\frac{d^{2}N}{dp_{T}dy} &= \frac{g e^{\frac{\mu}{T}}}{2\pi} p_{T}  m_{T} \frac{R}{\bar{v}_{T\mathrm{f}}} \int\limits_{0}^{R} r dr \int\limits_{-\eta_{max}}^{\eta_{max}} d\eta_{\|} \nonumber \\
                        & \times \cosh(\eta_{\|}-y) \ e^{-\frac{m_{T}}{T} \frac{\cosh(\eta_{\|}-y)}{\cosh\eta_{\|}}} \nonumber \\
                        & \times I_{0}\left(\frac{p_{T}}{T} \frac{\tanh\rho(r)}{\cosh\eta_{\|}} \right), \qquad \tanh\rho=\bar{v}_T.
\end{align}
Here, $T$ and $\mu$ represent the temperature and chemical potential of the fluid element in the laboratory reference frame $K$, respectively. Similarly, for the Maxwell-Boltzmann local equilibrium transverse momentum distribution in the Boltzmann-Gibbs blast-wave model using the Ott transformation ($l = 1$), we obtain
\begin{align}\label{t13}
&\frac{d^{2}N}{dp_{T}dy} = \frac{g e^{\frac{\mu}{T}}}{2\pi} p_{T}  m_{T} \frac{R}{\bar{v}_{T\mathrm{f}}} \int\limits_{0}^{R} r dr \int\limits_{-\eta_{max}}^{\eta_{max}} d\eta_{\|} \nonumber \\
                        & \times \cosh(\eta_{\|}-y) \ e^{-\frac{m_{T}}{T} \cosh^{2}\rho(r)\cosh\eta_{\|}\cosh(\eta_{\|}-y)} \nonumber \\
                        & \times I_{0}\left(\frac{p_{T}}{T} \cosh\rho(r)\sinh\rho(r)\cosh\eta_{\|} \right).
\end{align}
It should be noted that the transverse momentum distribution (\ref{t13}), derived using the Ott transformation, differs from the transverse momentum distribution (\ref{t12}), which is obtained through the Planck transformation.

\subsection{Tsallis blast-wave model for finite-volume fireballs}
The Tsallis blast-wave model is formulated based on the principle of local equilibrium, using Tsallis statistics~\cite{Tsal88,Tsal98} rather than the conventional Boltzmann-Gibbs statistics. By incorporating the Maxwell - Boltzmann single-particle distribution function in the zeroth-order approximation from the Tsallis-3 statistics, as outlined in Ref.~\cite{Parvan2023}, we develop the Tsallis blast-wave model for the Maxwell-Boltzmann statistics of particles in the zeroth-order approximation. This leads to the following expression for the Maxwell-Boltzmann local equilibrium transverse momentum distribution in the Tsallis blast-wave model (cf. Eq.~(\ref{t3}))
\begin{align}\label{t14}
\frac{d^{2}N}{dp_{T}dy} &= \frac{g}{(2\pi)^{3}} p_{T} \int d\Sigma_{\mu} p^{\mu} \nonumber \\
& \times  \int\limits_{0}^{2\pi} d\varphi \ \left[1-(1-q) \frac{p_{\mu}u^{\mu}-\mu_{0}}{T_{0}}\right]^{\frac{q}{1-q}},
\end{align}
where $q$ denotes the non-extensivity parameter. In the Gibbs limit $q\to 1$, the Tsallis blast-wave distribution (\ref{t14}) recovers the Boltzmann-Gibbs blast-wave distribution (\ref{t3}).

Using a cylindrically symmetric hypersurface and applying the calculation method outlined in~\cite{Florkowski}, we derive the Maxwell-Boltzmann local equilibrium transverse momentum distribution for the Tsallis blast-wave model in the zeroth-order approximation for a finite-volume freeze-out firecylinder, given by:
\begin{widetext}
\begin{align}\label{t14a}
\frac{d^{2}N}{dp_{T}dy} &= \frac{g}{(2\pi)^{3}} \ p_{T}  m_{T} \tau_{\mathrm{f}} \int\limits_{0}^{R} r dr \int\limits_{-\eta_{max}}^{\eta_{max}} d\eta_{\|} \cosh(\eta_{\|}-y) \int\limits_{0}^{2\pi} d\phi \int\limits_{0}^{2\pi} d\varphi  \nonumber \\
& \times  \left[1-(1-q) \frac{1}{T_{0}}\left\{m_{T}\cosh\rho(r)\cosh(\eta_{\|}-y) - p_{T} \sinh\rho(r)\cos(\phi-\varphi) - \mu_{0}\right\}\right]^{\frac{q}{1-q}}.
\end{align}
%\end{widetext}
Here, $\tau_{\mathrm{f}}=R/\bar{v}_{T\mathrm{f}}$, with $\cosh\rho$ and $\sinh\rho$ determined by Eqs.~(\ref{t5}) and (\ref{t6}). For a finite-volume freeze-out firecylinder with finite $\eta_{\text{max}}$, the transverse momentum distributions in Eqs.~(\ref{t3a}) and (\ref{t14a}) are not boost-invariant. These distributions correspond to finite volume fireballs produced in heavy-ion collisions. In contrast, boost-invariant distributions, commonly used in calculations~\cite{Florkowski}, originate from infinitely long freeze-out firecylinders with $\eta_{\text{max}}=\infty$ (see Eq.~(\ref{t8})), which are not realistic in heavy-ion collision processes. By integrating Eq.~(\ref{t14a}) over the azimuthal angles $\phi$ and $\varphi$ and applying the second relation from Eq.~(\ref{t6}), we obtain:
%\begin{widetext}
\begin{align}\label{t15}
\frac{d^{2}N}{dp_{T}dy} &= \frac{g}{2\pi} p_{T}  m_{T} \frac{R}{\bar{v}_{T\mathrm{f}}} \int\limits_{0}^{R} r dr \int\limits_{-\eta_{max}}^{\eta_{max}} d\eta_{\|} \cosh(\eta_{\|}-y) \left[1-(1-q) \frac{1}{T_{0}}\left(m_{T}\cosh\rho(r)\cosh(\eta_{\|}-y)-\mu_{0}\right)\right]^{\frac{q}{1-q}}  \nonumber \\
                        & \times \ _{2}F_{1}\left(\frac{1}{2}\frac{q}{q-1},\frac{1}{2}\frac{q}{q-1}+\frac{1}{2};1;\left[\frac{(1-q)\frac{p_{T}}{T_{0}} \sinh\rho(r)}{1-(1-q) \frac{1}{T_{0}}\left(m_{T}\cosh\rho(r)\cosh(\eta_{\|}-y)-\mu_{0}\right)}\right]^{2}   \right),
\end{align}
\end{widetext}
where $\ _{2}F_{1}(a,b;c;z)$ denotes the hypergeometric function. Here, $T_0$ and $\mu_0$ represent the rest temperature and rest chemical potential, respectively, of the fluid element in the reference frame $K_0$. Several studies~\cite{Tang2009,Shao2010,Waqas2020,Gu2022,Chen2021} utilizing the Tsallis blast-wave model exhibit flaws in their transverse momentum distributions. These works omit the factor $\tau_{\mathrm{f}}$, which depends on the transverse velocity $\bar{v}_{T\mathrm{f}}$. This simplification may lead to inaccurate numerical results. Additionally, these studies rely on an inconsistent single-particle Tsallis distribution (see the proof in Ref.~\cite{Parvan2021}). An exception is the study~\cite{Waqas2020}, which adopts the single-particle distribution of the Tsallis-3 statistics in the zeroth-order approximation. However, the authors of Ref.~\cite{Waqas2020} overlook the integration over the spacetime rapidity of the fluid element and assume $\eta_{||}=0$, which contradicts the blast-wave model's definition.

By applying the Planck transformations (\ref{39b}) and the Ott transformations (\ref{39bb}), we can rewrite the transverse momentum distribution (\ref{t14}) for the Tsallis blast-wave model as follows:
\begin{align}\label{t16}
\frac{d^{2}N}{dp_{T}dy} &= \frac{g}{(2\pi)^{3}} p_{T} \int d\Sigma_{\mu} p^{\mu} \nonumber \\
& \times  \int\limits_{0}^{2\pi} d\varphi \ \left[1-(1-q) \frac{\gamma^{l}p_{\mu}u^{\mu}-\mu}{T}\right]^{\frac{q}{1-q}},
\end{align}
where $T$ and $\mu$ denote the temperature and chemical potential, respectively, of the fluid element in the laboratory reference frame $K$.

By substituting Eq.~(\ref{t11}) into Eq.~(\ref{t16}) and applying the same derivation method used for Eq.~(\ref{t15}), we derive the Maxwell-Boltzmann transverse momentum distribution for the Tsallis blast-wave model, incorporating the Planck transformations with $l=-1$, as follows:
\begin{widetext}
\begin{align}\label{t17}
\frac{d^{2}N}{dp_{T}dy} &= \frac{g}{2\pi} p_{T}  m_{T} \frac{R}{\bar{v}_{T\mathrm{f}}} \int\limits_{0}^{R} r dr \int\limits_{-\eta_{max}}^{\eta_{max}} d\eta_{\|} \cosh(\eta_{\|}-y) \left[1-(1-q) \frac{1}{T}\left(\frac{m_{T}\cosh(\eta_{\|}-y)}{\cosh\eta_{\|}}-\mu\right)\right]^{\frac{q}{1-q}}  \nonumber \\
                        & \times \ _{2}F_{1}\left(\frac{1}{2}\frac{q}{q-1},\frac{1}{2}\frac{q}{q-1}+\frac{1}{2};1;\left[\frac{(1-q)\frac{p_{T}}{T}\frac{\tanh\rho(r)}{\cosh\eta_{\|}} }{1-(1-q) \frac{1}{T}\left(\frac{m_{T}\cosh(\eta_{\|}-y)}{\cosh\eta_{\|}}-\mu\right)}\right]^{2}   \right),
\end{align}
%\end{widetext}
Similarly, for the Maxwell-Boltzmann transverse momentum distribution of the Tsallis blast-wave model using the Ott transformations with $l = 1$, we obtain:
%\begin{widetext}
\begin{align}\label{t18}
\frac{d^{2}N}{dp_{T}dy} &= \frac{g}{2\pi} p_{T}  m_{T} \frac{R}{\bar{v}_{T\mathrm{f}}} \int\limits_{0}^{R} r dr   \nonumber \\  & \times
\int\limits_{-\eta_{max}}^{\eta_{max}} d\eta_{\|} \cosh(\eta_{\|}-y) \left[1-(1-q) \frac{1}{T}\left(m_{T}\cosh^{2}\rho(r)\cosh\eta_{\|}\cosh(\eta_{\|}-y)-\mu\right)\right]^{\frac{q}{1-q}}  \nonumber \\
                        & \times \ _{2}F_{1}\left(\frac{1}{2}\frac{q}{q-1},\frac{1}{2}\frac{q}{q-1}+\frac{1}{2};1;\left[\frac{(1-q)\frac{p_{T}}{T}\cosh\rho(r) \sinh\rho(r)\cosh\eta_{\|}}{1-(1-q) \frac{1}{T}\left(m_{T}\cosh^{2}\rho(r)\cosh\eta_{\|}\cosh(\eta_{\|}-y)-\mu\right)}\right]^{2}   \right).
\end{align}
\end{widetext}
It should be noted that the transverse momentum distribution (\ref{t18}), derived using the Ott transformation, differs from the transverse momentum distribution (\ref{t17}), which is obtained through the Planck transformation. Note that studies on the Tsallis thermometer are explored in Ref.~\cite{Biro3}.

\subsection{Numerical results} In the Maxwell-Boltzmann global equilibrium transverse momentum distributions of the Boltzmann-Gibbs statistics (\ref{t1}) and the Tsallis-3 statistics in the zeroth-order approximation (\ref{t2}), the temperature $T$ and chemical potential $\mu$ are defined for the globally equilibrated fireball in the laboratory reference frame $K$. In contrast, the blast-wave model employs the rest temperature $T_0$ and rest chemical potential $\mu_0$ of the fluid element, which remains stationary in the moving reference frame $K_0$. Consequently, the temperature and chemical potential differ between the distributions of the global equilibrium statistical model and the local equilibrium blast-wave model, as they are defined in distinct reference frames.

\begin{figure}[!htb]
\includegraphics[width=0.49\textwidth]{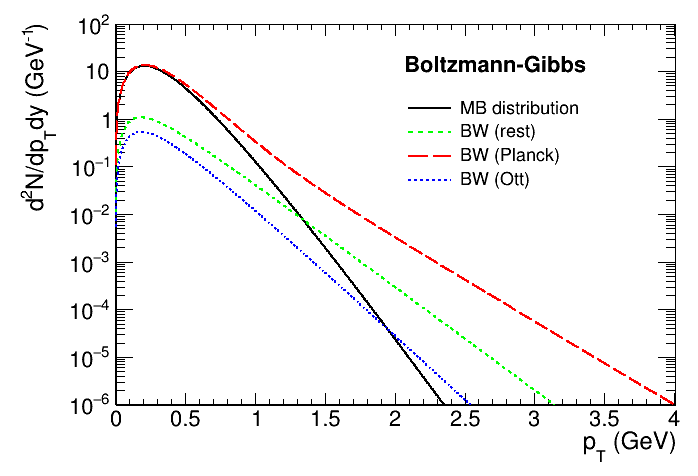}
\caption{(Color online) Comparison of the Maxwell-Boltzmann global equilibrium transverse momentum distribution (solid line) for $\pi^{+}$ pions with the Maxwell-Boltzmann local equilibrium transverse momentum distributions from the blast-wave (BW) model, using Planck transformations (long dashed line) and Ott transformations (dotted line). The distributions are evaluated at a temperature of $T=100$ MeV and a chemical potential of $\mu=10$ MeV in the laboratory reference frame $K$ at a rapidity of $y=0$. Calculations are based on a firecylinder at freeze-out with a radius of $R=2$ fm and a maximum spacetime rapidity of $\eta_{\text{max}}=3$. For the BW model, a transverse flow velocity of $\bar{v}_{T\mathrm{f}}=0.6$ is used. Transverse momentum distribution (\ref{t4}) with rest-frame parameters (dashed line), defined in the rest frame of a fluid element, are evaluated at $T_0 = 100$ MeV and $\mu_0 = 10$ MeV.}
\label{fig3}
\end{figure}

By aligning the temperature and chemical potential of these two approaches in the same reference frame, comparing their distributions can determine whether the Planck or Ott transformations produce consistent results. We expect that transforming the temperature and chemical potential in the $ p_T $-distributions of the blast-wave model to the laboratory frame $ K $ would produce a quantitative alignment with standard Maxwell-Boltzmann global equilibrium $ p_T $-distributions. For the Planck transformations, we have the Maxwell - Boltzmann local equilibrium transverse momentum distribution (\ref{t12}) for the Boltzmann-Gibbs blast-wave model and the local equilibrium distribution (\ref{t17}) for the Tsallis blast-wave model. For the Ott transformations, we have the Maxwell-Boltzmann local equilibrium transverse momentum distribution (\ref{t13}) for the Boltzmann-Gibbs blast-wave model and the local equilibrium distribution (\ref{t18}) for the Tsallis blast-wave model. Thus, the Planck and Ott transformations produce distinct distributions in the blast-wave model when the temperature and chemical potential are considered in the laboratory reference frame $K$.

Figure~\ref{fig3} illustrates the Maxwell-Boltzmann global equilibrium transverse momentum distribution (solid line), alongside the Maxwell-Boltzmann local equilibrium transverse momentum distributions of the Boltzmann - Gibbs blast-wave model with the rest-frame parameters (dashed line), and using Planck's transformation (long dashed line) and Ott's transformation (dotted line). These distributions were calculated for $\pi^{+}$ pions using Eqs.~(\ref{t1}), (\ref{t4}), (\ref{t12}), and (\ref{t13}), respectively, with a temperature of $T = 100$ MeV and a chemical potential of $\mu = 10$ MeV, defined in the laboratory reference frame $K$ at a rapidity of $y = 0$. The calculations assume a firecylinder at freeze-out with a radius of $R = 2$ fm and a maximum spacetime rapidity of $\eta_{max} = 3$. For the blast-wave model, a transverse flow velocity of $\bar{v}_{T\mathrm{f}} = 0.6$ was adopted. For the distribution (\ref{t4}), we have taken $T_0 = 100$ MeV and $\mu_0 = 10$ MeV.

\begin{figure}[!htb]
\includegraphics[width=0.49\textwidth]{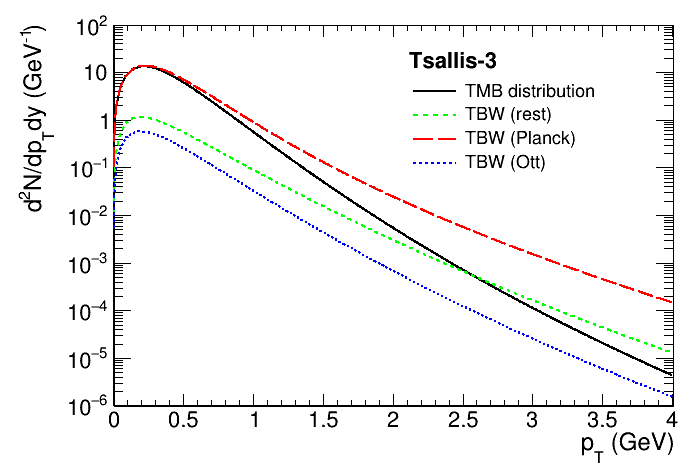}
\caption{(Color online) Comparison of the Maxwell-Boltzmann global equilibrium transverse momentum distribution of the Tsallis-3 statistics in the zeroth-order approximation (solid line) for $\pi^{+}$ pions with the Maxwell-Boltzmann local equilibrium transverse momentum distributions of the Tsallis blast-wave (TBW) model with rest-frame parameters (dashed line) and using Planck transformations (long dashed line) and Ott transformations (dotted line), at a non-extensivity parameter of $q=1.05$. Further details are provided in the caption of Figure~\ref{fig3}.}
\label{fig4}
\end{figure}

The local equilibrium Boltzmann-Gibbs blast-wave model using Planck transformations closely matches the Maxwell-Boltzmann global equilibrium transverse momentum distribution at their peak values, indicating strong compatibility. In contrast, the peak value of the same model with Ott transformations is significantly lower than those of the Maxwell-Boltzmann global equilibrium distribution and the blast-wave model with rest-frame parameters. Thus, Planck transformations ensure consistency between global and local transverse momentum distributions within the Boltzmann-Gibbs statistical framework, whereas Ott transformations introduce notable discrepancies.

Figure~\ref{fig4} depicts the Maxwell-Boltzmann global equilibrium transverse momentum distribution of the Tsallis-3 statistics in the zeroth-order approximation (solid line), alongside the Maxwell-Boltzmann local equilibrium transverse momentum distributions of the Tsallis blast-wave model with rest-frame parameters (dashed line), and using Planck's transformation (long dashed line) and Ott's transformation (dotted line). These distributions were computed for $\pi^{+}$ pions using Eqs.~(\ref{t2}), (\ref{t15}), (\ref{t17}), and (\ref{t18}), with the same parameter values as those employed for the Boltzmann-Gibbs statistics in Figure~\ref{fig3}, and an additional non-extensivity parameter of $q = 1.05$.

The local equilibrium Tsallis blast-wave model with Planck's transformation closely aligns with the Maxwell-Boltzmann global equilibrium transverse momentum distribution of the Tsallis-3 statistics in the zeroth-order approximation at their peak values, indicating strong compatibility. In contrast, the peak value of the local equilibrium Tsallis blast-wave model with Ott's transformation is significantly lower than those of the Tsallis-3 global equilibrium distribution and the Tsallis blast-wave model with rest-frame parameters. Thus, Planck transformations ensure consistency between global and local transverse momentum distributions within Tsallis statistical framework, whereas Ott transformations introduce discrepancies.

The difference between distributions using Ott and Planck transformations stems from the Ott temperature and chemical potential deviating from their Planck counterparts by a factor of $\gamma^2$ (see Eq.~(\ref{s6})). The similarity between the $p_T$-distribution with Planck ($l=-1$) transformations (Eq.~(\ref{t10})) and the Maxwell-Boltzmann distribution (Eq.~(\ref{t1})) in the peak region at small $p_T$ arises because both distributions share the same particle energy term, $-E_{\mathbf{p}}/T$, in their exponents in the laboratory reference frame $K$. In contrast, the $p_T$-distribution with Ott ($l=1$) transformations (Eq.~(\ref{t10})) differs, as its exponent incorporates a distinct term, $-\gamma^2 E_{\mathbf{p}}/T$. This common dependence on $-E_{\mathbf{p}}/T$ in the Maxwell-Boltzmann and Planck-transformed $p_T$ - distributions underscores their compatibility, while the deviation in the Ott-transformed distribution suggests potential inaccuracies. The same reasoning extends to the Tsallis model, as outlined in Eqs.~(\ref{t2}) and (\ref{t16}).

\section{Conclusions}\label{sec8}
In this study, we investigate the Planck and Ott transformations of thermodynamic quantities, grounding our analysis in relativistic mechanics and equilibrium thermodynamics. We analyze a thermodynamic system stationary in reference frame $K_0$, which moves at constant velocity relative to the laboratory frame $K$. Our approach combines thermodynamic potentials with the principles of relativistic mechanics.

We employ a thermodynamic framework to derive relativistic transformations of thermodynamic quantities. For a thermodynamic system in relativistic motion within the laboratory reference frame $K$, we define two primary thermodynamic potentials from which all other potentials and ensembles are derived through Legendre transformations or by substituting the state variable, entropy. The first, the fundamental thermodynamic potential, represents the system's total relativistic energy (the Hamiltonian) and is expressed as a function of momentum as an independent state variable. The second, the conjugate fundamental thermodynamic potential, is the negative Lagrangian, expressed as a function of velocity as an independent state variable. Equilibrium thermodynamics is incorporated into relativistic dynamics through the rest energy of a thermodynamic system, which corresponds to the internal energy of the system and serves as the fundamental thermodynamic potential in the reference frame $K_0$, where the system is at rest. The fundamental thermodynamic potential and its conjugate are not related through Legendre transformations when substituting the variables of velocity and momentum. This is because the rest energy of a thermodynamic system, which serves as a thermodynamic potential, depends on the thermodynamic state variables. The thermodynamic potential obtained from the fundamental thermodynamic potential via Legendre transformations, with the momentum variable substituted by velocity, differs from the conjugate fundamental thermodynamic potential (the negative Lagrangian) by a factor of $\alpha$. Conversely, the thermodynamic potential derived from the conjugate fundamental thermodynamic potential through Legendre transformations, with the velocity variable replaced by momentum, differs from the fundamental thermodynamic potential (total energy) and precisely aligns with the formula for energy from Tolman.

In this paper, we derive relativistic transformations for thermodynamic quantities in the fundamental ensemble and its conjugate, ensuring consistency with the first law of thermodynamics (energy conservation) and the principle of entropy invariance. The fundamental thermodynamic potential, represented by the Hamiltonian and expressed as a function of momentum as an independent state variable, yields relativistic transformations for thermodynamic quantities that differ from the Planck transformations by a factor of $\alpha$. In contrast, the conjugate fundamental thermodynamic potential, embodied by the negative Lagrangian, produces the Planck transformations for thermodynamic quantities. The Ott transformations for thermodynamic quantities are derived from the system's total relativistic energy, expressed as a function of velocity as an independent state variable. However, using total energy in this way does not yield a thermodynamic potential and is inconsistent with the principles of Hamiltonian mechanics.

In this paper, we systematically explored the thermodynamic relationships of an ultrarelativistic ideal gas of quarks and gluons within the Stefan-Boltzmann limit. Using this model, we examined the temperature and pressure behavior in a relativistically moving quark-gluon plasma system, analyzing their dependence on the system's momentum. We determined that the Ott trace anomaly of an ideal gas aligns with the Stefan – Boltzmann equation, whereas the Planck trace anomaly does not.

We have developed and formulated the consistent Boltzmann-Gibbs and Tsallis blast-wave models, defined within the finite volume of the freeze-out firecylinder. Our results reveal that the commonly used formula for the transverse momentum distribution in the Boltzmann-Gibbs blast-wave model assumes an infinite freeze-out firecylinder volume--an uncommon scenario in heavy ion collisions. Furthermore, we formulated Maxwell-Boltzmann transverse momentum distributions for the Boltzmann-Gibbs and Tsallis blast-wave models in finite volume systems, incorporating Planck and Ott transformations to fix the temperature and chemical potential in the laboratory reference frame $K$. By comparing the local equilibrium transverse momentum distributions of these models, which include Planck and Ott transformations, with the global equilibrium distributions derived from conventional Boltzmann - Gibbs and Tsallis statistics in the same reference frame, we found that only Planck transformations yield consistent results, whereas Ott transformations result in significant discrepancies.

\vskip0.2in
\noindent
{\bf Data availability}

No data was used for the research described in the article.

\vskip0.1in
\noindent
{\bf Acknowledgments}

This work was supported in part by the RSCF grant, N22-72-10028, N22-72-10028-P and the Romanian Ministry of Research, Innovation and Digitalization, Project PN 23 21 01 01/2023. We thank A.A. Aparin and E.V. Nedorezov for discussions on the subject matter of this work.

\appendix{}

%\vskip0.2in

\section{Derivation of Planck and Ott Transformations}\label{App}
In special relativity, for a system moving at a constant velocity $\mathbf{v} = \text{const}$ with $d\mathbf{v} = 0$, the Lorentz transformation for the infinitesimal change in energy between two inertial reference frames, $K$ and $K_0$, can be expressed as
\begin{equation}\label{1ap}
\mathrm{d}E_0 = \gamma \left( \mathrm{d}E - \mathbf{v} \cdot \mathrm{d}\mathbf{P} \right),
\end{equation}
where $\gamma = (1 - \mathbf{v})^{-1/2}$ is the Lorentz factor, $\mathrm{d}E$ is the energy differential in frame $K$, $\mathrm{d}E_0$ is the energy differential in frame $K_0$, and $\mathrm{d}\mathbf{P}$ is the momentum differential in frame $K$. The following relations hold for infinitesimal changes in entropy ($S$), volume ($V$), and momentum ($\mathbf{P}$):
\begin{equation}\label{2ap}
\mathrm{d}S_0 = \mathrm{d}S, \qquad \mathrm{d}V_0 = \gamma \mathrm{d}V, \qquad \mathrm{d}\mathbf{P} = \mathbf{v} \mathrm{d}E.
\end{equation}
These relations account for the transformation properties of thermodynamic quantities under a Lorentz boost.

\paragraph{Ott Transformations}
The temperature $T_0$ in the rest frame $K_0$ of a thermodynamic system is defined as the ratio of energy to entropy differentials:
\begin{align}\label{3ap}
T_0 &= \frac{\mathrm{d}E_0}{\mathrm{d}S_0} = \frac{\gamma \left( \mathrm{d}E - \mathbf{v} \cdot \mathrm{d}\mathbf{P} \right)}{\mathrm{d}S} \nonumber \\
&= \gamma \left( \frac{\mathrm{d}E}{\mathrm{d}S} - \mathbf{v} \cdot \frac{\mathrm{d}\mathbf{P}}{\mathrm{d}S} \right).
\end{align}
Substituting $\mathrm{d}\mathbf{P} = \mathbf{v} \mathrm{d}E$ from Eq.~(\ref{2ap}), we obtain:
\begin{align}\label{3app}
T_0 &=  \gamma \frac{\mathrm{d}E}{\mathrm{d}S} \left( 1 - \mathbf{v}^{2} \right)  = \frac{1}{\gamma} \frac{\mathrm{d}E}{\mathrm{d}S} = \frac{T_{\text{Ott}}}{\gamma},
\end{align}
where the temperature in frame $K$ is defined as:
\begin{equation}\label{4ap}
 T_{\text{Ott}} \equiv \frac{\mathrm{d}E}{\mathrm{d}S}.
\end{equation}
Equation~(\ref{3app}) is the Ott transformation for temperature.

Similarly, the pressure $p_0$ in frame $K_0$ is defined as:
\begin{align}\label{5ap}
p_0 &= -\frac{\mathrm{d}E_0}{\mathrm{d}V_0} = -\frac{\gamma \left( \mathrm{d}E - \mathbf{v} \cdot \mathrm{d}\mathbf{P} \right)}{\gamma \mathrm{d}V} \nonumber \\ &= -\left( \frac{\mathrm{d}E}{\mathrm{d}V} - \mathbf{v} \cdot \frac{\mathrm{d}\mathbf{P}}{\mathrm{d}V} \right).
\end{align}
Using $\mathrm{d}\mathbf{P} = \mathbf{v} \mathrm{d}E $, this becomes:
\begin{equation}\label{5app}
p_0 = -\frac{\mathrm{d}E}{\mathrm{d}V} \left( 1 - \mathbf{v}^{2} \right) = -\frac{\mathrm{d}E}{\mathrm{d}V} \cdot \frac{1}{\gamma^2} = \frac{p_{\text{Ott}}}{\gamma^2},
\end{equation}
where the pressure in frame $K$ is defined as:
\begin{equation}\label{6ap}
 p_{\text{Ott}} \equiv -\frac{\mathrm{d}E}{\mathrm{d}V}.
\end{equation}
Equation~(\ref{5app}) is the Ott transformation for pressure.

\paragraph{Planck transformations}
For the Planck transformations, the temperature $T_0$ in frame $K_0$ is similarly defined:
\begin{align}\label{7ap}
T_0 &= \frac{\mathrm{d}E_0}{\mathrm{d}S_0} =  \gamma \frac{\mathrm{d}E - \mathbf{v} \cdot \mathrm{d}\mathbf{P}}{\mathrm{d}S} = \gamma T_{\text{Pl}},
\end{align}
where the temperature in frame $K$ is defined as:
\begin{equation}\label{8ap}
  T_{\text{Pl}} \equiv \frac{\mathrm{d}E - \mathbf{v} \cdot \mathrm{d}\mathbf{P}}{\mathrm{d}S} = \frac{\mathrm{d} \bar{L}}{\mathrm{d} S}.
\end{equation}
Here, $\mathrm{d} \bar{L} = \mathrm{d}E - \mathbf{v} \cdot \mathrm{d}\mathbf{P}$ is the change in the negative Lagrangian. Equation~(\ref{7ap}) represents the Planck transformation for temperature. Substituting $\mathrm{d}\mathbf{P} = \mathbf{v} \mathrm{d}E$ into Eq.~(\ref{8ap}), we find:
\begin{equation}\label{9ap}
T_{\text{Pl}} = \frac{\mathrm{d}E}{\mathrm{d}S} \left( 1 - \mathbf{v}^{2} \right) = \frac{\mathrm{d}E}{\mathrm{d}S} \cdot \frac{1}{\gamma^2} = \frac{T_{\text{Ott}}}{\gamma^2},
\end{equation}
using the definition of $T_{\text{Ott}}$ from Eq.~(\ref{4ap}).

The pressure $p_0$ in frame $K_0$ is given by:
\begin{align}\label{10ap}
p_0 &= -\frac{\mathrm{d}E_0}{\mathrm{d}V_0} =  -\frac{\mathrm{d}E - \mathbf{v} \cdot \mathrm{d}\mathbf{P}}{\mathrm{d}V} = p_{\text{Pl}},
\end{align}
where the pressure in frame $K$ is defined as:
\begin{equation}\label{11ap}
p_{\text{Pl}} \equiv -\frac{\mathrm{d}E - \mathbf{v} \cdot \mathrm{d}\mathbf{P}}{\mathrm{d}V}  = - \frac{\mathrm{d} \bar{L}}{\mathrm{d} V}.
\end{equation}
Equation~(\ref{10ap}) is the Planck transformation for pressure. Substituting $\mathrm{d}\mathbf{P} = \mathbf{v} \mathrm{d}E $ into Eq.~(\ref{11ap}), we obtain:
\begin{align}\label{12ap}
p_{\text{Pl}} &=  -\frac{\mathrm{d}E}{\mathrm{d}V} \left( 1 - \mathbf{v}^{2} \right) = -\frac{\mathrm{d}E}{\mathrm{d}V} \cdot \frac{1}{\gamma^2} = \frac{p_{\text{Ott}}}{\gamma^2},
\end{align}
using the definition of $p_{\text{Ott}}$ from Eq.~(\ref{6ap}).


\begin{thebibliography}{100}
\bibitem{Haar} D.~ter~Haar, H.~Wergeland, Phys. Rep. 1C (1971) 31.

\bibitem{Farias} C.~Farias, V.A.~Pinto, P.S.~Moya, Sci. Rep. 7 (2017) 17657.

\bibitem{Nakamura} T.K.~Nakamura, Prog. Theor. Phys. 128 (2012) 463.


\bibitem{Mosengeil} K.~von~Mosengeil, Annalen der Physik 327, 5 (1907) 867-904.

\bibitem{Planck} M.~Planck, Annalen der Physik 331, 6 (1908) 1-34; Sitzungsberichte der K\"{o}niglich Preussischen Akademie der Wissenschaften, June 13, 1907, pp.~542-570.

\bibitem{Einstein} A.~Einstein, Jahrbuch der Radioaktivit\"{a}t und Elektronik 4 (1907) 411-462.

\bibitem{Ott} H.~Ott, Z. Physik 175 (1963) 70-104.


\bibitem{Arzelies} H.~Arzeli\`{e}s, Nuovo Cimento B 40 (1965) 333.

\bibitem{Kibble} T.W.B.~Kibble, Nuovo Cimento B 41 (1966) 72.

\bibitem{Sutcliffe} W.G.~Sutcliffe, Nuovo Cimento 39 (1965) 683.

\bibitem{Landsberg} P.T.~Landsberg, Nature 212 (1966) 571.

\bibitem{Cavalleri} G.~Cavalleri, G.~Salgarelli, Nuovo Cimento A 62 (1969) 722.

\bibitem{Newburgh} R.G.~Newburgh, Nuovo Cimento B 52 (1979) 219.

\bibitem{Yuen} C.K.~Yuen,  Am. J. Phys. 38 (1970) 246.

\bibitem{Callen} H.~Callen, G.~Horwitz, Am. J. Phys. 39 (1971) 938.

\bibitem{Treder} H.-J.~Treder, Annalen der Physik 489, 1 (1977) 23-29.

\bibitem{Pathria} R.K.~Pathria, Proc. Phys. Soc. 88 (1966) 791.

\bibitem{Fenech} Ch.~Fenech, J.P.~Vigier, Phys. Lett. A 215 (1996) 247.

\bibitem{Kampen} N.G.~van~Kampen, Phys. Rev. 173 (1968) 295.

\bibitem{Israel} W.~Israel, Annals of Physics 100 (1976) 310.

\bibitem{Landsberg1996} P.T~Landsberg, G.E.A.~Matsas, Phys. Lett. A 223 (1996) 401.

\bibitem{Mares} J.J.~Mare\v{s},  P.~Hub\'{\i}k, J.~\v{S}est\'{a}k, V.~\v{S}pi\v{c}ka, J.~Kri\v{s}tofik, J.~St\'{a}vek, Physica E 42 (2010) 484.

\bibitem{Papadatos} N.~Papadatos, C.~Anastopoulos, Phys. Rev. D 102 (2020) 085005.

\bibitem{Parga} G.~Ares~de~Parga, B.~L\'{o}pez-Carrera, F.~Angulo-Brown, J. Phys. A: Math. Gen. 38 (2005) 2821.

\bibitem{Heras} J.A.~Heras, M.G.~Osorno, Eur. Phys. J. Plus 137 (2022) 423.

\bibitem{Mares2017} J.J.~Mare\v{s},  P.~Hub\'{\i}k, V.~\v{S}pi\v{c}ka, Fortschr. Phys. 65 (2017) 1700018.

\bibitem{Mi} D.~Mi, H.Y.~Zhong, D.M.~Tong, Mod. Phys. Lett. A 24 (2009) 73.

\bibitem{Agmon} N.~Agmon, Foundations of Physics 7 (1977) 331.

\bibitem{Balescu} R.~Balescu, Physica 40 (1968) 309.

\bibitem{Guemez} J.~G\"{u}\'{e}mez, J.A.~Mier, Phys. Scr. 98 (2023) 025001.

\bibitem{Gavassino} L.~Gavassino, Foundations of Physics 50 (2020) 1554.

\bibitem{Przanowski} M.~Przanowski, J.~Tosiek, Phys. Scr. 84 (2011) 055008.

\bibitem{Montakhab} A.~Montakhab, M.~Ghodrat, M.~Barati, Phys. Rev. E 79 (2009) 031124.

\bibitem{Biro} T.S.~Bir\'{o}, P.~V\'{a}n, Europhys. Lett. 89 (2010) 30001.

\bibitem{Hao} X.~Hao, S.~Liu, L.~Zhao, Commun. Theor. Phys. 75 (2023) 035601.

\bibitem{Parvan2019} A.S.~Parvan, Annals of Physics 401 (2019) 130.



\bibitem{Biro1} T.S.~Bir\'{o}, Is There a Temperature?: Conceptual Challenges at High Energy, Acceleration and Complexity, Springer, 2011.

\bibitem{Becattini1}  F. Becattini, M. Buzzegoli, A. Palermo,  Phys. Lett. B 820 (2021) 136519.

\bibitem{Becattini2} F. Becattini, L. Bucciantini, E. Grossi, and L. Tinti, Eur. Phys. J. C 75 (2015) 191.

\bibitem{Gavassino22} L.~Gavassino, Foundations of Physics 52 (2022) 11.


\bibitem{Landau} L.D.~Landau, E.M.~Lifshitz, The Classical Theory of Fields, Vol. 2, third ed., Pergamon Press, 1971.


\bibitem{Prigogine1} I.~Prigogine, D.~Kondepudi, Modern Thermodynamics: From Heat Engines to Dissipative Structures, John Wiley $\&$ Sons, 1998.

\bibitem{Kvasnikov} I.A.~Kvasnikov, Thermodynamics and Statistical Mechanics: The Equilibrium Theory, Moscow State Univ. Publ., 1991 (In Russian).

\bibitem{Prigogine2} I.~Prigogine, R.~Defay, Chemical Thermodynamics, Longmans, Green $\&$ Co Ltd, 1954.

\bibitem{Parvan15} A.S.~Parvan, Foundation of Equilibrium Statistical Mechanics Based on Generalized Entropy, Recent Advances in Thermo and Fluid Dynamics, Intech, Rijeka, 2015, Ed.~Mofid Gorji-Bandpy, Ch.11, http://dx.doi.org/10.5772/60997.


\bibitem{Tolman} R.C.~Tolman, Relativity, Thermodynamics and Cosmology, Oxford: Clarendon Press, 1934.


\bibitem{Baldin}  A.M.~Baldin, V.I.~Goldanskii, and I.L.~Rozenthal', Kinematics of Nuclear Reactions, Oxford University Press, London, 1961.


\bibitem{Halzen} F.~Halzen and A.D.~Martin, Quarks and Leptones: An Introductory Course in Modern Particle Physics, John Wiley $\&$ Sons, 1991.

\bibitem{Bourrely2002} C.~Bourrely, J.~Soffer, F.~Buccella, Eur. Phys. J. C 23 (2002) 487–501.

\bibitem{Soffer2019}  J.~Soffer, C.~Bourrely,  Nuclear Physics A 991 (2019) 121607.

\bibitem{Burkert2018}  V.D.~Burkert, L.~Elouadrhiri, and  F.X.~Girod, Nature 557 (2018) 396–399.

\bibitem{Mac1989} E.~Mac and E.~Ugaz, Z. Phys. C - Particles and Fields 43 (1989) 655-661.


\bibitem{Yagi2008} Kohsuke Yagi, Tetsuo Hatsuda, and Yasuo Miake, Quark-Gluon plasma: from big bang to little bang, Cambridge University Press, 2008.

\bibitem{Brown} G.E.~Brown, H.A.~Bethe, P.M.~Pizzochero, Phys. Lett. B 263 (1991) 337.

\bibitem{Parvan2020} A.S.~Parvan, Eur. Phys. J. A 56 (2020) 192.


\bibitem{Parvan2017} A.S.~Parvan, O.V.~Teryaev, and J.~Cleymans, Eur. Phys. J. A 53 (2017) 102.

\bibitem{Cleymans12a} J.~Cleymans, D.~Worku, J. Phys. G: Nucl. Part. Phys. 39 (2012) 025006.

\bibitem{Cleymans2012}  J.~Cleymans, D.~Worku, Eur. Phys. J. A 48 (2012) 160.

\bibitem{Parvan2023} A.S.~Parvan, J. Phys. G: Nucl. Part. Phys. 50 (2023) 125002.

\bibitem{Tsal88} C.~Tsallis, J. Stat. Phys. 52, (1988) 479.

\bibitem{Tsal98} C.~Tsallis, R.S.~Mendes, A.R.~Plastino, Physica A 261 (1998) 534.

\bibitem{Florkowski} W.~Florkowski, Phenomenology of Ultra-Relativistic Heavy-Ion Collisions, World Scientific, 2010.

\bibitem{Schnedermann} E.~Schnedermann, J.~Sollfrank, and U.W.~Heinz, Phys. Rev. C. 48 (1993) 2462-2475.


\bibitem{Tang2009}  Z.~Tang, Y.~Xu, L.~Ruan, G.~van~Buren, F.~Wang, and Z.~Xu, Phys. Rev. C. 79 (2009) 051901(R).

\bibitem{Shao2010} M.~Shao, L.~Yi, Z.~Tang, H.~Chen, C.~Li, and Z.~Xu,  J. Phys. G: Nucl. Part. Phys. 37 (2010) 085104.

\bibitem{Waqas2020} M.~Waqas, F.-H. Liu, R.-Q. Wang, and I.~Siddique, Eur. Phys. J. A 56 (2020) 188.

\bibitem{Gu2022} J.~Gu, C.~Li, Q.~Wang, W.~Zhang, and H.~Zheng, J. Phys. G: Nucl. Part. Phys. 49 (2022) 115101.

\bibitem{Chen2021}  J.~Chen, J.~Deng, Z.~Tang, Z.~Xu, and L.~Yi, Phys. Rev. C. 104 (2021) 034901.



\bibitem{Parvan2021}  A.S.~Parvan and T.~Bhattacharyya, J. Phys. A: Math. Theor. 54 (2021) 325004.

\bibitem{Biro3} G.~B\'{i}r\'{o}, G.G.~Barnaf\"{o}ldi and T.S.~Bir\'{o},  J. Phys. G: Nucl. Part. Phys. 47 (2020) 105002.



\end{thebibliography}
\end{document}